\documentclass[11pt, a4paper, logo]{googlecloud}

\pdftrailerid{redacted}

\makeatletter
\renewcommand\bibentry[1]{\nocitep{#1}{\frenchspacing\@nameuse{BR@r@#1\@extra@b@citeb}}}
\makeatother

\usepackage{amsmath,amsfonts,bm}

\def\eqref#1{equation~\ref{#1}}

\def\1{\bm{1}}

\DeclareMathAlphabet{\mathsfit}{\encodingdefault}{\sfdefault}{m}{sl}
\SetMathAlphabet{\mathsfit}{bold}{\encodingdefault}{\sfdefault}{bx}{n}

\newcommand{\ours}{{\textsc{ScholarPeer}}\xspace}
\newcommand{\ourss}{{\textsc{ScholarPeer}}\xspace}
\newcommand{\backbone}{Gemini 3.1 Pro\xspace}

\newcommand{\hmaxmetric}{H-Max}

\definecolor{pastelgreen}{rgb}{0.88, 1.0, 0.88}
\definecolor{pastelred}{rgb}{1.0, 0.88, 0.88}
\definecolor{ARCBlue}{RGB}{20, 50, 100}

\usepackage{dsfont}

\usepackage[authoryear, sort&compress, round]{natbib}

\usepackage[utf8]{inputenc} %
\usepackage[T1]{fontenc}    %

\usepackage{microtype}
\usepackage{inconsolata}

\usepackage{booktabs}
\usepackage{graphicx}
\usepackage[most]{tcolorbox}
\usepackage{stackengine}
\usepackage{multirow}
\usepackage{subcaption} 
\usepackage{wrapfig}
\usepackage{hyperref}
\usepackage{makecell}
\usepackage{setspace}
\usepackage{booktabs} 
\usepackage{array}    
\usepackage{geometry} 

\usepackage{url}            
\usepackage{fontawesome5}
\usepackage{amsfonts}       
\usepackage{nicefrac}       
\usepackage{microtype}      
\usepackage{xcolor,colortbl}         
\usepackage{times}
\usepackage{multicol}
\usepackage{blindtext}
\usepackage{tabu}
\usepackage{amsmath, bm}

\usepackage{caption}
\usepackage[normalem]{ulem}
\usepackage{soul}
\usepackage{ragged2e}
\usepackage{pgffor}
\usepackage{textcomp}
\usepackage{amssymb}
\usepackage{pifont}
\usepackage{booktabs}
\usepackage{tabularx}
\usepackage{algorithm}
\usepackage{algpseudocode}
\usepackage{geometry}
\usepackage{booktabs} 
\usepackage{multirow} 
\usepackage{graphicx} 
\usepackage{paracol}
\usepackage{tcolorbox}
\tcbuselibrary{breakable}

\title{\ours~: A Multi-Agent Framework for Automated Peer Review}

\correspondingauthor{Palash Goyal \href{mailto:palashgoyal@agoogle.com}{<palashgoyal@google.com>}, and Jinsung Yoon \href{mailto:jinsungyoon@google.com}{<jinsungyoon@google.com>}}

\renewcommand{\today}{}

\author[1]{\fontsize{10.0pt}{10.0pt}\selectfont Palash Goyal}
\author[1]{\fontsize{10.0pt}{10.0pt}\selectfont Mihir Parmar}
\author[1]{\fontsize{10.0pt}{10.0pt}\selectfont Yiwen Song}
\author[1]{\fontsize{10.0pt}{10.0pt}\selectfont Hamid Palangi}
\author[1]{\fontsize{10.0pt}{10.0pt}\selectfont Tomas Pfister}
\author[1]{\fontsize{10.0pt}{10.0pt}\selectfont Jinsung Yoon}

\affil[1]{\fontsize{9.0pt}{9.0pt}\selectfont Google}

\begin{abstract}
    The exponential growth of machine learning submissions has strained the traditional peer review process, resulting in slow feedback loops for authors and an immense burden on reviewers to rigorously audit technical soundness and verify literature. To address this, we introduce \ours~, a multi-agent framework designed to operationalize the rigorous auditing workflow of a senior researcher. Rather than attempting to replace human judgment, \ours~ serves as a co-scientist: acting as a mentor for rapid author iteration prior to submission, and as an active verification assistant that augments human reviewers. The framework structurally decouples contextualization from critique by deploying a sub-domain historian to synthesize the field's trajectory, a baseline scout to proactively hunt for omitted state-of-the-art comparisons, and a multi-aspect Q\&A engine that deeply audits technical soundness—scrutinizing internal logical consistency, experimental validity, and mathematical rigor—while cross-referencing claims against top-tier academic venues. We comprehensively evaluate \ours~ on $\sim$1,800 ICLR submissions spanning 2020 through 2025. Our results show that \ours~ achieves significant win-rates against state-of-the-art fine-tuned models and search-augmented agentic baselines.
\end{abstract}

\begin{document}

\maketitle

\section{Introduction}
The democratization of artificial intelligence research has precipitated an unprecedented explosion in scientific output. Major machine learning conferences now receive tens of thousands of submissions annually, straining the peer review process to its breaking point. This scalability crisis manifests in two distinct operational bottlenecks: authors wait months to receive actionable feedback on their manuscripts, while area chairs and reviewers face extreme fatigue attempting to rigorously audit technical soundness, manually hunt down missing baselines, and verify state-of-the-art claims in a rapidly evolving literature space \citep{zhuang2025large, liang2024can}.

Consequently, the community has turned to Large Language Models (LLMs) to assist in the evaluation of scientific manuscripts \citep{luo2025llm4sr}. Early approaches primarily focused on fine-tuning models on static datasets of paper-review pairs \citep{liu2023reviewergpt, weng2024cycleresearcher}. While these systems excel at linguistic fluency and summarization, they evaluate papers in a parametric vacuum. To mitigate this, recent frameworks have integrated retrieval mechanisms \citep{zhu2025deepreview, zeng2025reviewrl, chitale2025autorev} or explored multi-agent review architectures \citep{jin2024agentreview, yamada2025ai}. However, these systems predominantly focus on optimizing text-generation policies or simulating collaborative discussions. They still lack the structural mechanisms necessary to conduct the rigorous auditing required to produce a publication-ready critique.

To address this gap, we introduce \ours~, a multi-agent framework  designed to operationalize the \textit{rigorous auditing} workflow of a senior researcher. Crucially, \ours~ is not designed to replace the human peer review system; rather, it serves as a co-scientist to augment it. For authors, \ours~ acts as a mentor, providing highly specific, actionable feedback that enables rapid iteration prior to submission. For reviewers and ACs, it serves as a verification assistant, auditing experimental validity, actively retrieving missing baselines, and contextualizing novelty to supply human experts with verifiable evidence, freeing them to focus on high-level scientific judgment.

Our framework structurally decouples paper comprehension from external contextualization. \ours~ employs three key agents: a \textit{sub-domain historian}, a \textit{baseline scout}, and a \textit{multi-aspect Q\&A engine}. The historian agent dynamically synthesizes a ``domain narrative,'' placing the submission in the context of the field's trajectory. The baseline scout proactively hunts for missing SOTA baselines and datasets that the authors failed to compare against. Finally, the Q\&A engine operates as a skeptic, deeply auditing technical soundness---scrutinizing internal logical consistency, experimental validity, and mathematical rigor---while cross-referencing claims against top-tier academic venues.

\begin{figure*}[ht]
    \centering
    \includegraphics[width=\linewidth]{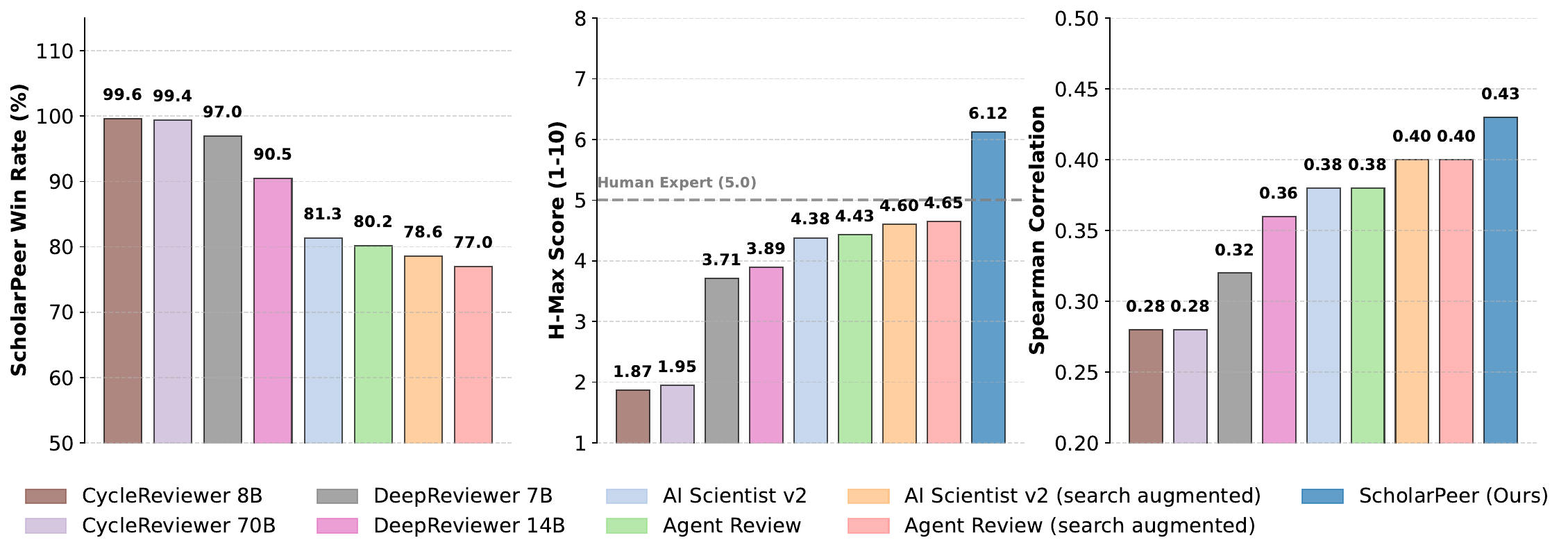}
    \caption{Comparative evaluation of \ours~ against existing frameworks on ScholarEval. (Left) Win rate of \ours~ against review fine-tuned models and agentic baselines. (Middle) Average \hmaxmetric~score (higher the better) of reviews generated by various frameworks (best human review is considered as 5). (Right) Spearman correlation of scores generated by review frameworks with ground-truth human rankings. We use Gemini 3.1 Pro as the backbone model for \ours~ and baseline agentic frameworks. We use Claude Sonnet 4.5 as the LLM-judge. These results show that our proposed framework consistently outperforms baselines across various metrics.}
    \label{fig:overall}
\end{figure*}

We comprehensively evaluate \ours~ on $\sim$1,800 submissions from ICLR 2020 through 2025, merging the evaluation sets of DeepReview-Bench\citep{zhu2025deepreview} and AgentReview \citep{jin2024agentreview}. We call this data set ScholarEval. Recognizing that traditional metrics often fail to capture the nuance of scientific critique, we also explore two new evaluation metrics: the \textit{H-Max score}, which calibrates the localized quality of critiques against human experts, and the \textit{review diversity score}, which assesses the variance of perspectives provided by the system.

\begin{figure*}[ht]
  \vskip 0.2in
  \begin{center}
    \centerline{\includegraphics[width=\textwidth]{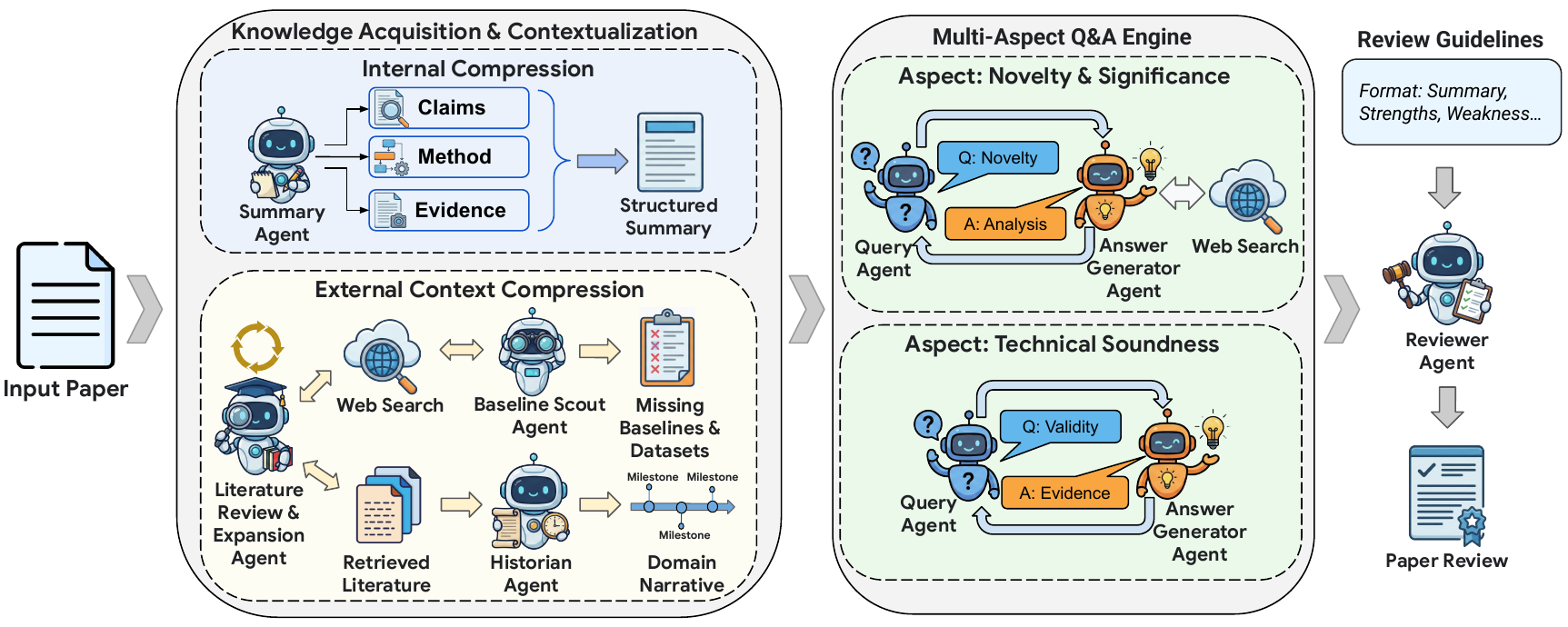}}
    \caption{
      The \ours~ framework: Given an input paper, the framework employs a dual-stream information retrieval process. The \textit{knowledge acquisition and contextualization module} uses summarizer, search-enabled literature review, historian and baseline scout  agents to compress internal and external information. These inputs feed into the \textit{multi-aspect Q\&A engine}, which generates and answers probing questions regarding the novelty and technical soundness. Finally, the \textit{review generator} utilizes these inputs and conference-specific review guidelines to generate the final review.
    }
    \label{icml-historical}
  \end{center}
  \vspace{-10mm}
\end{figure*}

Our main contributions are as follows:
\begin{itemize}
    \item We propose \ours~, a multi-agent framework that performs active, search-grounded interrogation, designed to accelerate author iteration and augment human reviewer judgment.
    \item We introduce specialized \textit{historian}, \textit{baseline scout}, and \textit{Q\&A} agents that collectively perform both internal logical consistency checks and external web-scale verification against top-tier academic venues.
    \item We conduct comprehensive evaluations, achieving dominant win-rates against state-of-the-art fine-tuned models and search-enabled agentic baselines, with practical deployment metrics ($\sim$\$1.20 and latency of $\sim$10 minutes per review).
\end{itemize}

\section{Problem Formulation}
\label{sec:problem}
We formulate the task of automated peer review assistance as generating a rigorous, verifiable critique $R$ given a paper submission $S$. To capture the cognitive process of a human expert, we formally distinguish between the intrinsic claims and experimental setup of the paper, $S_{content}$, and the extrinsic, ever-evolving context of the scientific field, $\mathcal{C}_{dynamic}$. A human reviewer does not evaluate $S_{content}$ in isolation; rather, they adopt a skeptical stance, stress-testing the methodology against a dynamic mental graph of prior art, concurrent work, and established community standards. Consequently, a rigorous automated review must actively interrogate $S_{content}$ while leveraging $\mathcal{C}_{dynamic}$, formulated as the mapping:
\begin{equation}
    R = f(S_{content}, \mathcal{C}_{dynamic}).
\end{equation}
We design \ours~ to execute this mapping through a multi-agent orchestration that dynamically constructs $\mathcal{C}_{dynamic}$, structurally transforms $S_{content}$, and actively interrogates the submission. Ultimately, this framework produces a strong, verifiable peer review designed to augment human judgment and accelerate author iteration.

\section{\ours~}
\label{sec:method}
In this section, we present the architecture of \ours~, an agentic framework for automated peer review. As illustrated in Figure \ref{icml-historical}, \ours~ operates via a dual-stream process designed to emulate the information-gathering and auditing flow of a human expert. Rather than processing the manuscript in a single pass---which often leads to the ``lost-in-the-middle'' phenomenon and surface-level summaries---the framework is organized into two primary subsystems: the Knowledge Acquisition \& Contextualization Module (\textsection\ref{sec:method_kac}) and the Multi-Aspect Q\&A Engine (\textsection\ref{sec:method_qa}).

\subsection{Knowledge Acquisition and Contextualization}
\label{sec:method_kac}
The knowledge acquisition module structurally transforms $S_{content}$ and dynamically constructs $\mathcal{C}_{dynamic}$ using four specialized agents.

\textbf{Summary Agent (Internal Extraction).} A significant bottleneck in applying LLMs to full-text papers is the ``lost-in-the-middle'' phenomenon and the cognitive overload associated with processing dense technical tokens simultaneously. To mitigate cognitive overload, the summary agent transforms the raw text of the submission into a structured representation, $\hat{S}$, extracting: (1) the core claim set $\mathcal{H}_{core}$, (2) the proposed method $M$, and (3) the reported evidence $E$. By decoupling structural comprehension from critique, this guides the downstream agents to operate on a high-fidelity signal of the paper's exact premises.

\textbf{Literature Review \& Expansion Agent (Dynamic Context Creation).} This agent constructs a ``live'' reference frame, enabling the system to validate claims against the latest literature. It constructs $\mathcal{C}_{dynamic}$ by executing a two-step retrieval process. First, it identifies the paper's sub-domain based on the abstract and performs an initial literature search using a search engine. Second, it iteratively identifies gaps in the literature search and performs an ``expansion search'' targeting recent pre-prints and concurrent work. Crucially, to ensure academic rigor and prevent low-quality source pollution, source quality filters strictly restrict search results to top-tier conferences (e.g., NeurIPS, ICLR, ICML, CVPR, ACL) and arXiv, enforcing a strict publication cutoff date matching the paper's submission. By relying exclusively on the Search API to retrieve actual, verifiable literature rather than generating URLs from parametric memory, this agent structurally minimizes hallucination.

\textbf{Sub-Domain Historian Agent (External Contextualization).} This agent augments $\mathcal{C}_{dynamic}$ with domain trajectory. Retrieving raw abstracts is insufficient to assess significance. The historian agent compresses the retrieved literature into a chronological ``domain narrative'', identifying the arc of progress in the sub-domain. This narrative mimics a senior researcher's mental model, enabling the system to evaluate the significance of the paper's contributions.

\textbf{Baseline Scout Agent (Adversarial Auditing).} This agent augments $\mathcal{C}_{dynamic}$ with independent baseline scouting. It acts as an adversarial auditor, independently searching for the current SOTA methods on the specific benchmarks used in the paper. It returns a concrete list of omitted competitors and datasets, providing the actionable evidence required for a rigorous technical soundness critique.

\subsection{Multi-Aspect Q\&A Engine and Review Generation}
\label{sec:method_qa}
The Multi-Aspect Q\&A Engine utilizes the structured and dynamic context to rigorously probe the paper. Finally, the review generator agent produces the final review using these signals.

\textbf{Multi-Aspect Q\&A Agents (Adversarial Probing).} We deploy dedicated query and answer generation agents for two critical aspects: \textit{novelty \& significance} and \textit{technical soundness}. The query agents act as ``skeptics'' to actively probe the paper, generating questions using the structured summary, the domain narrative, and the baseline scout's findings. The novelty \& significance query agent enumerates the distinct contribution claims and produces one question per claim. The technical soundness query agent generates questions to scrutinize the soundness of paper's critical aspects. To answer the novelty \& significance questions, the answer agent is provided with active search access. Conversely, the technical soundness answer agent cross-references the extracted claims against the paper's own methodology and evidence sets to find logical inconsistencies, impossible mathematical constraints, or inappropriate evaluation metrics (e.g., catching discretization errors in continuous treatment functions).

\textbf{Review Generator Agent (Guidelines-Driven Synthesis)}
The review generator agent synthesizes the final report by integrating the structured paper summary $\hat{S}$ and the verified Q\&A pairs from the interrogation log. This agent is conditioned on explicit \textit{review guidelines} (e.g., the NeurIPS reviewer checklist). By decoupling the ``investigation'' (Q\&A Engine) from the ``reporting'' (Review Generator), \ours~ effectively aggregates findings, resolves any minor query overlaps, and outputs a highly constructive, venue-aligned critique ready for author iteration or AC augmentation.
\section{Experiments}
\label{sec:experiments}
\subsection{Evaluation Protocols}
Evaluating automated critiques is challenging, as standard NLP metrics fail to capture scientific depth, and LLM judges can exhibit verbosity or family-model biases. Since our objective is to accelerate author iteration and augment human reviewers, our evaluation protocols prioritize the intrinsic quality, actionability, and depth of the generated critiques over predicting a paper's final acceptance score. To ensure rigorous, multi-faceted assessment, we triangulate our results using side-by-side evaluations, two novel metrics (\hmaxmetric~and review diversity), and decision alignment, validated against an expert human baseline. We provide the complete prompts used for evaluation in Appendix~\ref{app:eval_prompts}.

\textbf{Side-by-Side (SxS) Evaluation.} The search-enabled judge is provided with the input paper and reviews from two anonymized frameworks (with order randomized to avoid position bias). The judge compares the reviews across five dimensions: Technical Accuracy, Constructive Value, Analytical Depth, Significance Assessment, and Overall Judgment. These dimensions directly measure the framework's utility for authors and reviewers, strictly penalizing hallucinations and generic advice.

\textbf{\hmaxmetric~Score.} To measure how well an AI assistant actually augments the human review process, we introduce the \hmaxmetric~score. An expert judge (search-enabled) evaluates a single AI review for a given paper against the \textit{collective set} of human reviews for that paper.

The judge identifies the strongest points made by any human reviewer on a specific aspect (e.g., Technical Accuracy) and sets that as the ``Expert Baseline''. The AI review is then scored relative to this baseline. An \hmaxmetric~score of 5 signifies that the AI review is of similar quality to the strongest points made by a set of $k$ reviewers. A score of 10 signifies the AI review is transformative compared to the collective set of human reviews, and a score of 1 means the AI review misses critical points. Appendix~\ref{app:scoring} contains the complete mapping from score to its interpretation.

\textbf{Review Diversity Score (RDS).} A known systemic risk of LLM automation is the ``artificial hivemind''~\cite{jiang2025artificial} where models converge on homogenized opinions. To quantify perspective variance, we generate $N=3$ reviews for the same paper, compute embeddings using an embedding model (we used DistilRoberta-v1 \citep{sanh2019distilbert} for our experiments), and calculate the Inter-review Semantic Similarity ($IR_{sim}$) in embedding space:
$$IR_{sim} = \frac{1}{N(N-1)} \sum_{i \neq j} \text{CosineSim}(E(r_i), E(r_j))$$
We define the Review Diversity Score as $1 - IR_{sim}$. A higher RDS indicates the framework explores varied, rigorous perspectives rather than collapsing to a mean response.

\textbf{Decision Score Alignment.} While our primary goal is generating actionable critiques rather than predicting outcomes, prior automated review systems heavily rely on decision prediction. For completeness and comparability, we evaluate the alignment of the model's quantitative recommendations with human ground truth by reporting the Spearman Correlation ($\rho$) between the model's predicted final decision scores and the actual human rankings.

\subsection{Experimental Setup}
\begin{table*}[t]
\centering
\caption{Direct comparison of \ours~ (\backbone) with baselines across five critical dimensions on ScholarEval. ``Win'' indicates that the judge assessed \ours~ as superior (average over three runs). Cells highlighted in green denote the superior outcome of our framework (red indicates that the baseline is superior). \ours~ achieves dominant win-rates in \textit{Significance Assessment} and \textit{Constructive Value}, validating the impact of the multi-agent reasoning architecture. We use Claude Sonnet 4.5 as the LLM judge. We provide the standard deviation in bracket.}
\label{tab:sxs_fine_grained}
\resizebox{\textwidth}{!}{
\begin{tabular}{llcccccccccc}
\toprule
\multirow{2}{*}{\textbf{Category}} & \multirow{2}{*}{\textbf{Baselines}} & \multicolumn{2}{c}{\textbf{Technical Accuracy}} & \multicolumn{2}{c}{\textbf{Constructive Value}} & \multicolumn{2}{c}{\textbf{Analytical Depth}} & \multicolumn{2}{c}{\textbf{Significance Assessment}} & \multicolumn{2}{c}{\textbf{Overall Judgment}} \\
\cmidrule(lr){3-4} \cmidrule(lr){5-6} \cmidrule(lr){7-8} \cmidrule(lr){9-10} \cmidrule(lr){11-12}
 & & \textbf{Win(\%)$\uparrow$} & \textbf{Lose(\%)} & \textbf{Win(\%)$\uparrow$} & \textbf{Lose(\%)} & \textbf{Win(\%)$\uparrow$} & \textbf{Lose(\%)} & \textbf{Win(\%)$\uparrow$} & \textbf{Lose(\%)} & \textbf{Win(\%)$\uparrow$} & \textbf{Lose(\%)} \\
\midrule
\multirow{4}{*}{\textbf{Fine-tuned}} 
 & CycleReviewer-8B & \cellcolor{pastelgreen}\bfseries 98.6 (0.1) & 1.0 (0.1) & \cellcolor{pastelgreen}\bfseries 100.0 (0.0) & 0.0 (0.0) & \cellcolor{pastelgreen}\bfseries 100.0 (0.0) & 0.0 (0.0) & \cellcolor{pastelgreen}\bfseries 98.4 (0.2) & 0.3 (0.1) & \cellcolor{pastelgreen}\bfseries 99.6 (0.1) & 0.3 (0.1) \\
 & CycleReviewer-70B & \cellcolor{pastelgreen}\bfseries 98.5 (0.2) & 1.2 (0.1) & \cellcolor{pastelgreen}\bfseries 99.8 (0.1) & 0.2 (0.1) & \cellcolor{pastelgreen}\bfseries 99.8 (0.1) & 0.0 (0.0) & \cellcolor{pastelgreen}\bfseries 98.5 (0.3) & 0.5 (0.1) & \cellcolor{pastelgreen}\bfseries 99.4 (0.1) & 0.4 (0.1) \\
 & DeepReviewer-7B & \cellcolor{pastelgreen}\bfseries 92.1 (0.3) & 5.4 (0.2) & \cellcolor{pastelgreen}\bfseries 97.2 (0.2) & 2.2 (0.2) & \cellcolor{pastelgreen}\bfseries 97.2 (0.3) & 1.6 (0.2) & \cellcolor{pastelgreen}\bfseries 95.6 (0.3) & 1.9 (0.2) & \cellcolor{pastelgreen}\bfseries 97.0 (0.3) & 2.4 (0.3) \\
 & DeepReviewer-14B & \cellcolor{pastelgreen}\bfseries 77.3 (0.6) & 12.7 (0.6) & \cellcolor{pastelgreen}\bfseries 88.3 (0.5) & 9.5 (0.5) & \cellcolor{pastelgreen}\bfseries 85.6 (0.5) & 11.9 (0.4) & \cellcolor{pastelgreen}\bfseries 89.1 (0.4) & 5.7 (0.3) & \cellcolor{pastelgreen}\bfseries 90.5 (0.2) & 8.8 (0.2) \\
\midrule
\multirow{3}{*}{\textbf{Single Agent}} 
 & Claude Sonnet 4.5 & \cellcolor{pastelgreen}\bfseries 42.9 (0.9) & 26.4 (0.7) & \cellcolor{pastelgreen}\bfseries 71.2 (0.8) & 23.3 (0.6) & \cellcolor{pastelgreen}\bfseries 51.4 (0.9) & 38.5 (0.8) & \cellcolor{pastelgreen}\bfseries 77.3 (0.7) & 13.8 (0.5) & \cellcolor{pastelgreen}\bfseries 74.1 (0.8) & 24.5 (0.7) \\
 & Gemini 3.1 Flash & \cellcolor{pastelgreen}\bfseries 39.5 (1.1) & 32.4 (0.9) & \cellcolor{pastelgreen}\bfseries 81.6 (0.7) & 16.7 (0.6) & \cellcolor{pastelgreen}\bfseries 74.9 (0.8) & 18.7 (0.7) & \cellcolor{pastelgreen}\bfseries 77.9 (0.8) & 13.0 (0.5) & \cellcolor{pastelgreen}\bfseries 78.3 (0.9) & 21.7 (0.8) \\
 & Gemini 3.1 Pro & \cellcolor{pastelgreen}\bfseries 40.2 (1.0) & 33.0 (0.8) & \cellcolor{pastelgreen}\bfseries 76.9 (0.9) & 18.1 (0.7) & \cellcolor{pastelgreen}\bfseries 69.1 (1.1) & 22.5 (0.8) & \cellcolor{pastelgreen}\bfseries 74.9 (0.9) & 11.8 (0.6) & \cellcolor{pastelgreen}\bfseries 73.0 (1.0) & 26.1 (0.9) \\
\midrule
\multirow{3}{*}{\shortstack[l]{\textbf{Single Agent} \\ \textbf{w/ Search}}} 
 & Claude Sonnet 4.5 & \cellcolor{pastelgreen}\bfseries 41.7 (0.8) & 27.4 (0.7) & \cellcolor{pastelgreen}\bfseries 71.1 (0.9) & 23.3 (0.7) & \cellcolor{pastelgreen}\bfseries 50.8 (1.0) & 39.5 (0.9) & \cellcolor{pastelgreen}\bfseries 75.2 (0.8) & 15.7 (0.6) & \cellcolor{pastelgreen}\bfseries 72.1 (0.9) & 26.4 (0.8) \\
 & Gemini 3.1 Flash & \cellcolor{pastelgreen}\bfseries 36.5 (1.2) & 35.4 (1.0) & \cellcolor{pastelgreen}\bfseries 78.6 (0.8) & 19.7 (0.7) & \cellcolor{pastelgreen}\bfseries 71.9 (0.9) & 21.7 (0.8) & \cellcolor{pastelgreen}\bfseries 74.9 (0.9) & 16.0 (0.7) & \cellcolor{pastelgreen}\bfseries 75.3 (1.0) & 24.7 (0.9) \\
 & Gemini 3.1 Pro & \cellcolor{pastelgreen}\bfseries 37.2 (1.1) & 36.0 (0.9) & \cellcolor{pastelgreen}\bfseries 73.9 (1.0) & 21.1 (0.8) & \cellcolor{pastelgreen}\bfseries 66.1 (1.2) & 25.5 (0.9) & \cellcolor{pastelgreen}\bfseries 71.9 (1.0) & 14.8 (0.7) & \cellcolor{pastelgreen}\bfseries 70.0 (1.1) & 29.1 (1.0) \\
\midrule
\multirow{7}{*}{\textbf{Multi Agent}} 
 & Agent Review (Claude 4.5) & \cellcolor{pastelgreen}\bfseries 35.1 (1.2) & 29.1 (1.1) & \cellcolor{pastelgreen}\bfseries 60.4 (1.1) & 35.5 (1.0) & 32.4 (1.2) & \cellcolor{pastelred}\bfseries 55.6 (1.2) & \cellcolor{pastelgreen}\bfseries 66.2 (1.1) & 19.1 (0.8) & \cellcolor{pastelgreen}\bfseries 56.8 (1.0) & 42.5 (1.1) \\
 & Agent Review (Gemini Flash) & \cellcolor{pastelgreen}\bfseries 47.8 (1.2) & 14.7 (1.0) & \cellcolor{pastelgreen}\bfseries 86.2 (0.9) & 11.4 (0.8) & \cellcolor{pastelgreen}\bfseries 79.4 (1.0) & 13.6 (0.9) & \cellcolor{pastelgreen}\bfseries 75.5 (0.9) & 8.4 (0.6) & \cellcolor{pastelgreen}\bfseries 83.2 (0.8) & 15.1 (0.7) \\
 & Agent Review (Gemini Pro) & \cellcolor{pastelgreen}\bfseries 46.3 (1.0) & 27.1 (0.8) & \cellcolor{pastelgreen}\bfseries 83.9 (0.7) & 12.6 (0.6) & \cellcolor{pastelgreen}\bfseries 80.7 (0.8) & 12.8 (0.7) & \cellcolor{pastelgreen}\bfseries 79.2 (0.8) & 10.0 (0.6) & \cellcolor{pastelgreen}\bfseries 80.2 (0.8) & 17.9 (0.6) \\
 & AI Scientist v2 (Claude 4.5) & \cellcolor{pastelgreen}\bfseries 37.2 (1.1) & 26.2 (0.9) & \cellcolor{pastelgreen}\bfseries 70.4 (1.0) & 23.6 (0.8) & 44.7 (1.2) & \cellcolor{pastelred}\bfseries 46.8 (1.0) & \cellcolor{pastelgreen}\bfseries 70.9 (0.9) & 19.4 (0.7) & \cellcolor{pastelgreen}\bfseries 64.7 (0.9) & 34.6 (0.8) \\
 & AI Scientist v2 (Gemini Flash) & \cellcolor{pastelgreen}\bfseries 48.1 (1.1) & 24.2 (0.9) & \cellcolor{pastelgreen}\bfseries 86.0 (0.9) & 10.4 (0.8) & \cellcolor{pastelgreen}\bfseries 76.7 (1.1) & 13.8 (0.9) & \cellcolor{pastelgreen}\bfseries 74.2 (1.0) & 11.5 (0.8) & \cellcolor{pastelgreen}\bfseries 81.1 (0.8) & 17.0 (0.7) \\
 & AI Scientist v2 (Gemini Pro) & \cellcolor{pastelgreen}\bfseries 49.1 (0.9) & 25.9 (0.7) & \cellcolor{pastelgreen}\bfseries 82.8 (0.6) & 14.1 (0.5) & \cellcolor{pastelgreen}\bfseries 82.0 (0.7) & 11.5 (0.6) & \cellcolor{pastelgreen}\bfseries 78.5 (0.7) & 10.9 (0.6) & \cellcolor{pastelgreen}\bfseries 81.3 (0.7) & 18.1 (0.6) \\
 & Stanford Agent Reviewer* & \cellcolor{pastelgreen}\bfseries 36.0 (1.4) & 32.0 (1.2) & \cellcolor{pastelgreen}\bfseries 52.0 (1.3) & 46.0 (1.1) & 38.0 (1.4) & \cellcolor{pastelred}\bfseries 46.0 (1.3) & \cellcolor{pastelgreen}\bfseries 60.0 (1.2) & 34.0 (1.0) & \cellcolor{pastelgreen}\bfseries 54.0 (1.1) & 46.0 (1.0) \\
\midrule
\multirow{6}{*}{\shortstack[l]{\textbf{Multi Agent} \\ \textbf{w/ Search}}} 
 & Agent Review (Claude 4.5) & \cellcolor{pastelgreen}\bfseries 30.5 (1.4) & 36.0 (1.2) & \cellcolor{pastelgreen}\bfseries 56.4 (1.3) & 41.6 (1.1) & 29.5 (1.5) & \cellcolor{pastelred}\bfseries 58.8 (1.3) & \cellcolor{pastelgreen}\bfseries 66.1 (1.2) & 19.2 (0.9) & \cellcolor{pastelgreen}\bfseries 54.9 (1.1) & 44.6 (1.2) \\
 & Agent Review (Gemini Flash) & \cellcolor{pastelgreen}\bfseries 44.7 (1.3) & 20.8 (1.1) & \cellcolor{pastelgreen}\bfseries 84.1 (1.0) & 13.3 (0.9) & \cellcolor{pastelgreen}\bfseries 74.6 (1.1) & 19.9 (1.0) & \cellcolor{pastelgreen}\bfseries 72.4 (1.0) & 16.4 (0.7) & \cellcolor{pastelgreen}\bfseries 79.5 (0.9) & 20.0 (0.8) \\
 & Agent Review (Gemini Pro) & \cellcolor{pastelgreen}\bfseries 42.3 (1.1) & 31.2 (0.9) & \cellcolor{pastelgreen}\bfseries 80.6 (0.8) & 16.0 (0.7) & \cellcolor{pastelgreen}\bfseries 77.1 (0.9) & 16.4 (0.8) & \cellcolor{pastelgreen}\bfseries 76.0 (0.9) & 13.1 (0.6) & \cellcolor{pastelgreen}\bfseries 77.0 (0.9) & 21.9 (0.7) \\
 & AI Scientist v2 (Claude 4.5) & \cellcolor{pastelgreen}\bfseries 34.2 (1.2) & 29.1 (1.0) & \cellcolor{pastelgreen}\bfseries 66.8 (1.1) & 29.6 (0.9) & 41.7 (1.3) & \cellcolor{pastelred}\bfseries 49.7 (1.1) & \cellcolor{pastelgreen}\bfseries 69.9 (1.0) & 20.1 (0.8) & \cellcolor{pastelgreen}\bfseries 62.8 (1.0) & 37.2 (0.9) \\
 & AI Scientist v2 (Gemini Flash) & \cellcolor{pastelgreen}\bfseries 45.3 (1.2) & 17.2 (1.0) & \cellcolor{pastelgreen}\bfseries 83.0 (1.0) & 13.4 (0.9) & \cellcolor{pastelgreen}\bfseries 70.7 (1.2) & 18.8 (1.0) & \cellcolor{pastelgreen}\bfseries 68.2 (1.1) & 18.5 (0.9) & \cellcolor{pastelgreen}\bfseries 76.0 (0.9) & 20.0 (0.8) \\
 & AI Scientist v2 (Gemini Pro) & \cellcolor{pastelgreen}\bfseries 46.2 (1.0) & 28.8 (0.8) & \cellcolor{pastelgreen}\bfseries 80.8 (0.7) & 16.2 (0.6) & \cellcolor{pastelgreen}\bfseries 79.3 (0.8) & 14.4 (0.7) & \cellcolor{pastelgreen}\bfseries 75.0 (0.8) & 13.7 (0.7) & \cellcolor{pastelgreen}\bfseries 78.6 (0.8) & 21.4 (0.7) \\
\bottomrule
\multicolumn{12}{l}{\textsuperscript{*}Stanford Agent Reviewer is not open-source and has only browser based access; evaluation was performed on 50 papers sampled from ScholarEval.}
\end{tabular}
}
\vspace{-5mm}
\end{table*}

\textbf{Dataset.} To rigorously evaluate our framework across a longitudinal time horizon, we curate ScholarEval. This dataset merges the test split of DeepReview-Bench \citep{zhu2025deepreview}, covering ICLR submissions from 2024 and 2025, with the evaluation set from AgentReview \citep{jin2024agentreview}, covering ICLR submissions from 2020 to 2023, resulting in a comprehensive benchmark of 1,786 machine learning submissions and their corresponding human reviews. By spanning a half-decade of research—including oral, spotlight, poster, and rejected papers—ScholarEval provides a highly diverse, real-world testbed to assess the temporal robustness of automated critiques.

\textbf{Baselines.} We categorize our baselines into three distinct groups to isolate the impact of architecture, training data, and retrieval capabilities: (1) Fine-tuned Baselines: We compare against CycleReviewer 8B and 70B \citep{weng2024cycleresearcher} and DeepReviewer 7B and 14B \citep{zhu2025deepreview}. These models represent the state-of-the-art in supervised fine-tuning on review data. Notably, DeepReviewer natively utilizes OpenScholar for dynamic literature search, providing a strong retrieval-enabled benchmark. (2) Agentic Baselines: We compare against single agent baselines and multi-agent baselines of AgentReview \citep{jin2024agentreview} and AI Scientist v2 \citep{yamada2025ai}, instantiated with various state-of-the-art backbone models including Gemini 3.1 Flash, Gemini 3.1 Pro and Claude Sonnet 4.5. (3) Search-Augmented Agentic Baselines: To address potential information asymmetry and prove that performance gains stem from our adversarial workflow, we equip the agentic baselines with the exact same Google Search tool used by ScholarPeer. In addition, we evaluate against Stanford Agent Reviewer (SAR)\footnote{\url{https://paperreview.ai/}}. Note that SAR is not open-source and has only browser based access. To evaluate SAR,
we sampled 50 papers from ScholarEval and got the reviews. 

\textbf{Implementation.} We use \backbone as the backbone LLM for all the agents in \ours for the main results and use Claude 4.5 Sonnet for the LLM judge. We further show results on backbone independence by swapping our backbone and the LLM judge. For web search, we use Google search and set publication cutoff date as the boundary for literature search. All automated metrics are averaged over three independent runs, with standard deviations reported to ensure statistical rigor. We provide further experimental details in Appendix~\ref{app:experiment_details} and the agent prompts in Appendix~\ref{app:agent_prompts}.

\textbf{LLM Judge Selection.} Before large-scale benchmarking, we validated the alignment between our Claude 4.5 Sonnet judge and human experts on a subset of 100 papers from ScholarEval. The LLM judge achieved a strong Pearson correlation of 0.53 with human experts and an inter-reviewer agreement of 88\% (Cohen's Kappa = 0.76). The guidelines provided to the expert researchers for performing the evaluation can be found in Appendix~\ref{app:human_eval_guidelines}. Further, a common issue of LLM-as-a-judge evaluations is a bias toward longer outputs. To ensure our win-rates reflect true critique quality rather than verbosity, we analyzed review lengths. \ours~ reviews averaged 597 words ($\sigma=61$). In contrast, human reviews averaged 788 words ($\sigma=295$), and DeepReviewer-14B averaged 2,156 words ($\sigma=312$). Despite generating significantly more concise reviews ($\sim$3.6$\times$ shorter than the strongest fine-tuned baseline and shorter than human experts), \ours~ consistently won SxS evaluations, showing that verbosity bias is not a concern for this task.

\subsection{Side-by-Side Results}
\begin{table*}[t]
\centering
\caption{\hmaxmetric~score, human correlation, and review diversity score on ScholarEval. \hmaxmetric~score shows comparison against best human expert reviews. Human corr ($\rho$) denotes Spearman correlation with human rankings. Review diversity measures semantic variance. \ours~ achieves state-of-the-art performance across all metrics. We provide the standard deviation over three runs in parentheses for the \hmaxmetric~scores.}
\label{tab:single_sided}
\resizebox{\textwidth}{!}{
\begin{tabular}{llccccccc}
\toprule
\multirow{2}{*}{\textbf{Category}} & \multirow{2}{*}{\textbf{Model}} & \multicolumn{5}{c}{\textbf{\hmaxmetric~Score Across Dimensions (1-10)$\uparrow$}} & \textbf{Human} & \textbf{Review} \\
\cmidrule(lr){3-7}
 & & \textbf{Tech. Acc.} & \textbf{Cons. Val.} & \textbf{Analytical} & \textbf{Signif.} & \textbf{Overall} & \textbf{Corr ($\rho$)$\uparrow$} & \textbf{Diversity$\uparrow$} \\
\midrule
\multirow{4}{*}{\textbf{Fine-tuned}} 
 & CycleReviewer-8B & 1.98 (0.01) & 2.17 (0.02) & 1.99 (0.01) & 2.31 (0.02) & 1.87 (0.01) & 0.28 & 0.01 \\
 & CycleReviewer-70B & 2.14 (0.02) & 2.31 (0.01) & 2.08 (0.02) & 2.64 (0.01) & 1.95 (0.02) & 0.28 & 0.01 \\
 & DeepReviewer-7B & 3.44 (0.01) & 3.55 (0.02) & 3.70 (0.01) & 3.65 (0.02) & 3.71 (0.01) & 0.32 & 0.02 \\
 & DeepReviewer-14B & 3.86 (0.02) & 3.98 (0.01) & 3.93 (0.02) & 3.80 (0.01) & 3.89 (0.02) & 0.36 & 0.02 \\
\midrule
\multirow{3}{*}{\textbf{Single Agent}} 
 & Claude Sonnet 4.5 & 4.69 (0.02) & 4.41 (0.01) & 4.63 (0.02) & 4.40 (0.01) & 4.52 (0.02) & 0.31 & 0.23 \\
 & Gemini 3.1 Flash & 4.58 (0.01) & 4.17 (0.02) & 4.46 (0.01) & 3.89 (0.02) & 4.34 (0.01) & 0.33 & 0.21 \\
 & Gemini 3.1 Pro & 4.86 (0.02) & 4.33 (0.01) & 4.53 (0.02) & 3.95 (0.01) & 4.50 (0.02) & 0.37 & 0.23 \\
\midrule
\multirow{3}{*}{\shortstack[l]{\textbf{Single Agent} \\ \textbf{w/ Search}}} 
 & Claude Sonnet 4.5 & 4.82 (0.02) & 4.55 (0.01) & 4.75 (0.02) & 4.90 (0.01) & 4.75 (0.01) & 0.31 & 0.23 \\
 & Gemini 3.1 Flash & 4.80 (0.01) & 4.35 (0.02) & 4.60 (0.01) & 4.45 (0.02) & 4.55 (0.02) & 0.36 & 0.25 \\
 & Gemini 3.1 Pro & 4.95 (0.02) & 4.50 (0.01) & 4.65 (0.02) & 4.55 (0.01) & 4.65 (0.02) & 0.39 & 0.25 \\
\midrule
\multirow{7}{*}{\textbf{Multi Agent}} 
 & Agent Review (Claude 4.5) & 4.59 (0.01) & 4.50 (0.02) & 4.72 (0.01) & 4.40 (0.02) & 4.58 (0.01) & 0.38 & 0.22 \\
 & Agent Review (Gemini Flash) & 4.41 (0.02) & 4.14 (0.01) & 4.24 (0.02) & 3.91 (0.01) & 4.19 (0.02) & 0.36 & 0.21 \\
 & Agent Review (Gemini Pro) & 4.77 (0.01) & 4.32 (0.02) & 4.40 (0.01) & 4.05 (0.02) & 4.43 (0.01) & 0.38 & 0.22 \\
 & AI Scientist v2 (Claude 4.5) & 4.66 (0.02) & 4.48 (0.01) & 4.70 (0.02) & 4.53 (0.01) & 4.60 (0.02) & 0.37 & 0.22 \\
 & AI Scientist v2 (Gemini Flash) & 4.71 (0.01) & 4.33 (0.02) & 4.48 (0.01) & 4.16 (0.02) & 4.36 (0.01) & 0.33 & 0.20 \\
 & AI Scientist v2 (Gemini Pro) & 4.63 (0.02) & 4.29 (0.01) & 4.35 (0.02) & 4.03 (0.01) & 4.38 (0.02) & 0.38 & 0.22 \\
\midrule
\multirow{6}{*}{\shortstack[l]{\textbf{Multi Agent} \\ \textbf{w/ Search}}} 
 & Agent Review (Claude 4.5) & 4.75 (0.01) & 4.65 (0.02) & 4.85 (0.01) & 4.95 (0.02) & 4.80 (0.01) & 0.39 & 0.22 \\
 & Agent Review (Gemini Flash) & 4.55 (0.02) & 4.35 (0.01) & 4.45 (0.02) & 4.40 (0.01) & 4.45 (0.02) & 0.38 & 0.24 \\
 & Agent Review (Gemini Pro) & 4.90 (0.01) & 4.50 (0.02) & 4.60 (0.01) & 4.60 (0.02) & 4.65 (0.01) & 0.40 & 0.23 \\
 & AI Scientist v2 (Claude 4.5) & 4.90 (0.02) & 4.65 (0.01) & 4.85 (0.02) & 5.10 (0.01) & 4.85 (0.02) & 0.38 & 0.22 \\
 & AI Scientist v2 (Gemini Flash) & 4.85 (0.01) & 4.60 (0.02) & 4.65 (0.01) & 4.60 (0.02) & 4.90 (0.01) & 0.36 & 0.23 \\
 & AI Scientist v2 (Gemini Pro) & 4.80 (0.02) & 4.45 (0.01) & 4.55 (0.02) & 4.45 (0.01) & 4.60 (0.02) & 0.39 & 0.23 \\
 & Stanford Agent Reviewer & 5.50 (0.03) & 5.39 (0.02) & \cellcolor{pastelgreen}\bfseries 6.22 (0.03) & 5.56 (0.02) & 5.89 (0.02) & -* & 0.26 \\
\midrule
\multirow{1}{*}{\textbf{Ours}} 
 & \ours & \cellcolor{pastelgreen}\bfseries 5.76 (0.01) & \cellcolor{pastelgreen}\bfseries 5.87 (0.02) & 5.85 (0.02) & \cellcolor{pastelgreen}\bfseries 6.47 (0.01) & \cellcolor{pastelgreen}\bfseries 6.12 (0.01) & \cellcolor{pastelgreen}\bfseries 0.43 & \cellcolor{pastelgreen}\bfseries 0.31 \\
\bottomrule
\multicolumn{9}{l}{\textsuperscript{*}Stanford Agent Reviewer does not provide a score.}
\end{tabular}
}
\vspace{-5mm}
\end{table*}
Table \ref{tab:sxs_fine_grained} illustrates the win rate of our framework against various baselines. \ours~ achieves a dominant 90.5\% win-rate against the strongest fine-tuned baseline, DeepReviewer-14B. Against the strongest agentic baseline with the same backbone model, we achieve a win rate of 73\%. Even against the strongest closed source multi-agent baseline, Stanford Agent Reviewer, we achieve a 54\% win-rate. Despite Claude Sonnet 4.5 being the judge, it scores ScholarPeer consistently higher than Claude based agentic baselines. This illustrates that our framework clearly outperforms the baselines.

\textbf{Addressing Information Asymmetry.} The single and multi-agent baselines do not leverage search tool. To further isolate our multi-agent architecture from information asymmetry, we evaluated \ours~ against \textit{search-augmented} variants of our agentic baselines. Even when equipped with the exact same Google Search tool, \ours~ outperforms AI Scientist v2 and AgentReview with search access by significant margins. This shows that the performance gains stem from the agentic architecture rather than search tool addition.

\textbf{Backbone Independence.} To confirm that our architecture—rather than the \backbone LLM—drives performance, we conducted a reverse-ablation using Claude 4.5 Sonnet as the backbone for \ours~ and the agentic baselines and using \backbone as the LLM judge. We achieve 90.2\% win rate against DeepReviewer-14B, 79.6\% against AI Scientist v2 and 78.8\% against AgentReview, confirming the framework's efficacy is model-agnostic.

\begin{figure*}[t]
    \centering
    \begin{subfigure}[b]{0.32\textwidth}
        \centering
        \includegraphics[width=\textwidth]{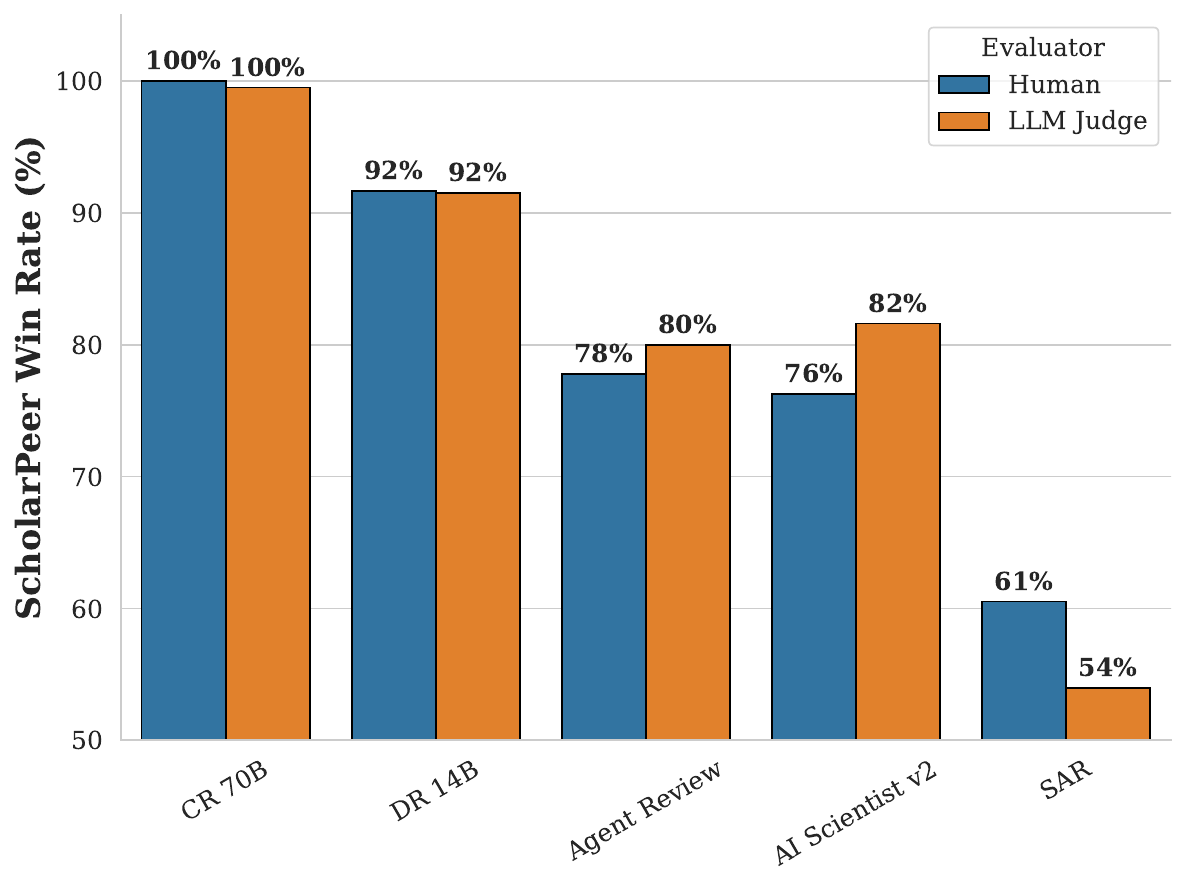}
        \caption{Side-by-Side (SxS) comparison of human vs. LLM judge win rates.}
        \label{fig:human_eval_sxs}
    \end{subfigure}
    \hfill
    \begin{subfigure}[b]{0.32\textwidth}
        \centering
        \includegraphics[width=\textwidth]{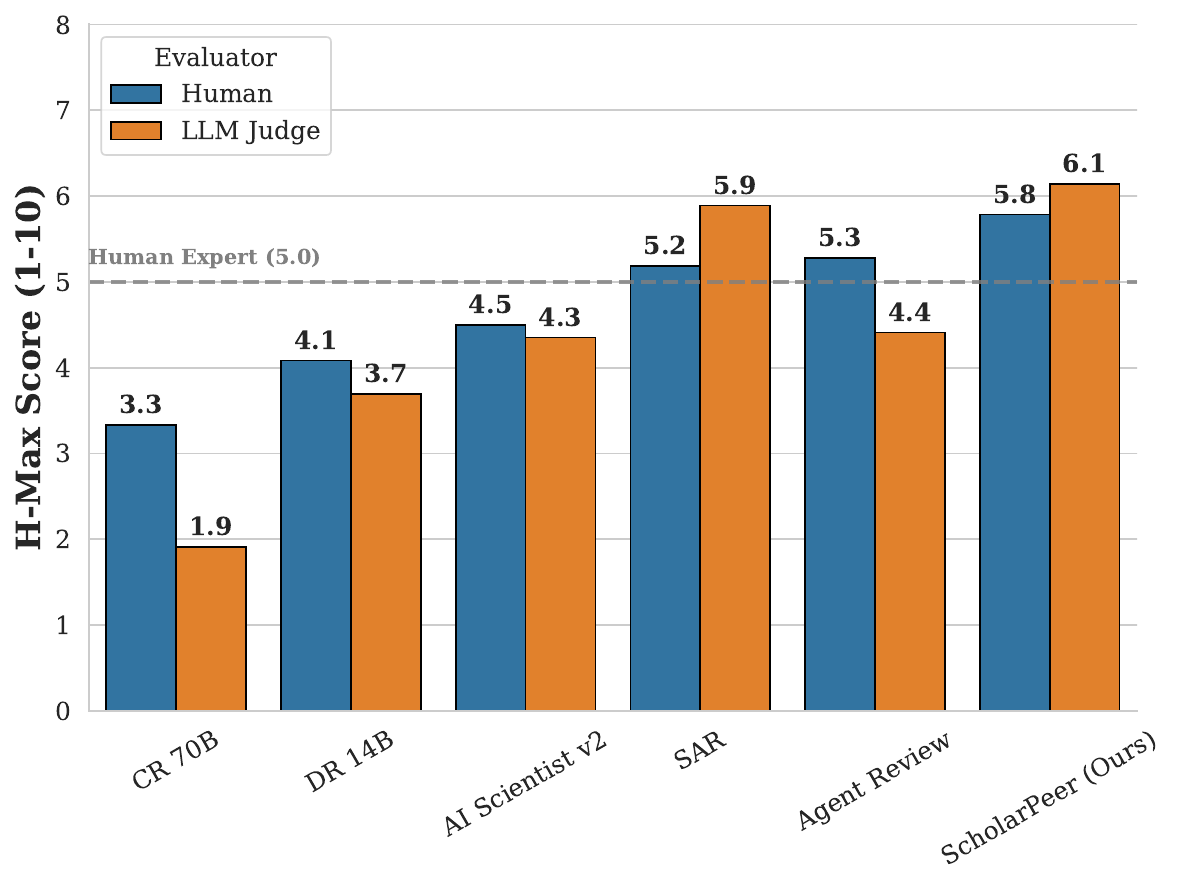}
        \caption{\hmaxmetric~score comparison of human evaluator vs LLM judge.}
        \label{fig:human_eval_single_sided}
    \end{subfigure}
    \hfill
    \begin{subfigure}[b]{0.32\textwidth}
        \centering
        \includegraphics[width=\textwidth]{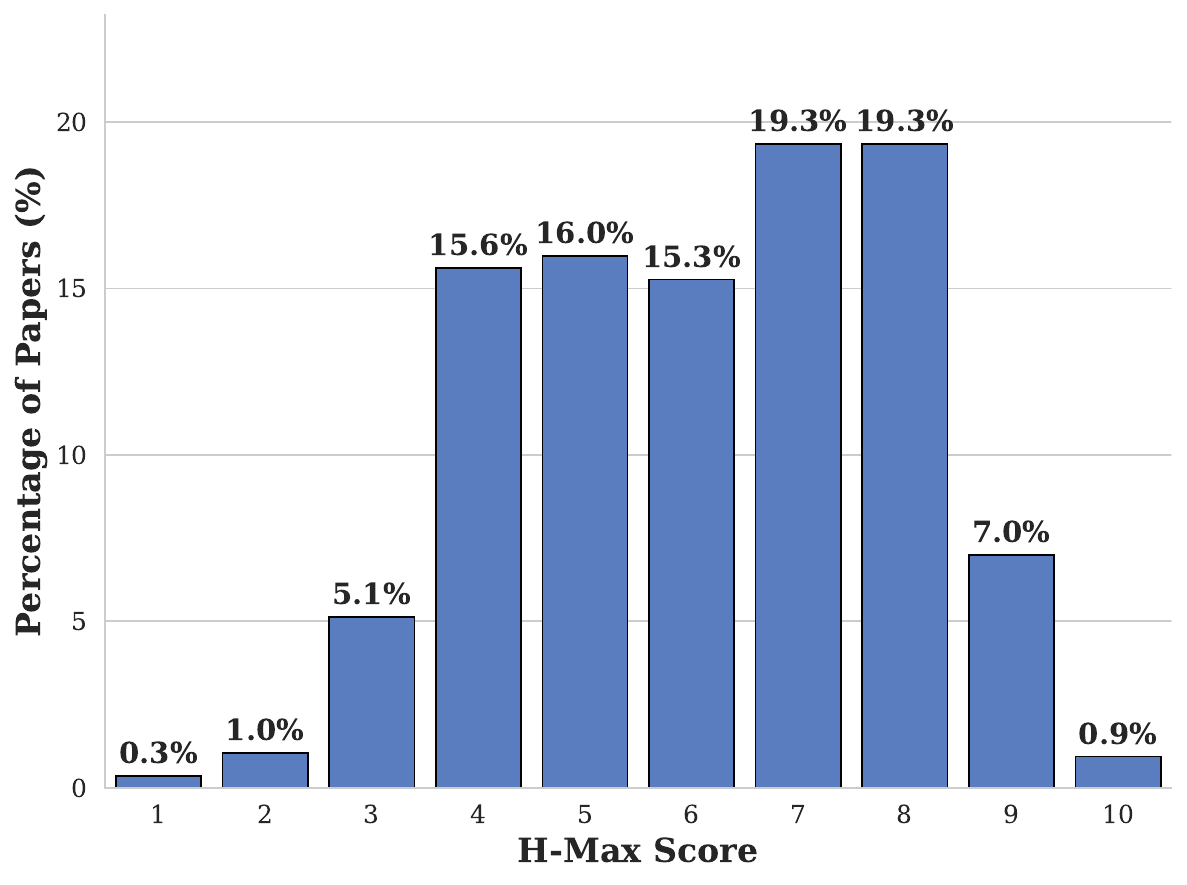}
        \caption{Distribution of overall scores for the \ourss framework.}
        \label{fig:overall_distribution}
    \end{subfigure}
    
    \caption{Alignment between LLM and human judges. We observe strong agreement in (a) side-by-side (SxS) win rates and (b) \hmaxmetric~score. Panel (c) illustrates that $>25\%$ of the reviews generated by \ourss achieve a score of 8 or higher.}
    \label{fig:combined_eval_results}
    \vspace{-5mm}
\end{figure*}

\textbf{Human Evaluation.} To complement our automated metrics, blinded expert reviewers compared \ours~ against five baselines (20 papers per baseline). Human experts consistently preferred \ours, with trends strongly correlating with the LLM judge, further validating our automated SxS protocol (see Figure \ref{fig:human_eval_sxs}).

\subsection{\hmaxmetric~Results}
Table \ref{tab:single_sided} presents the \hmaxmetric~scores. While state-of-the-art fine-tuned models struggle to surpass a score of 4.0, \ours~ achieves an exceptional average \hmaxmetric~score of 6.12. This score is driven by the dimensions most valuable for human augmentation. \ours~ scores an 6.47 in \textit{Significance Assessment} (contextualizing novelty) and 5.87 in \textit{Constructive Value} (providing exact, actionable remedies). Furthermore, \ours~ exhibits remarkable stability; across three independent runs, the overall \hmaxmetric~score yielded an exceptionally low standard deviation of $\sigma = 0.01$. Further, Figure~\ref{fig:overall_distribution} shows the distribution of score. We observe a significant $>25\%$ of generated reviews getting a score of 8.0 or above. 

\textbf{Human Evaluation.} Figure \ref{fig:human_eval_single_sided}
 shows that \ourss achieves the highest average rating across both human evaluation and LLM judge.

\subsection{Review Diversity}
As shown in Table \ref{tab:single_sided}, fine-tuned models suffer from severe mode collapse, exhibiting extremely low review diversity (RDS $\approx 0.01$--$0.02$) and homogenizing scientific feedback. Single and multi-agent baselines improve this (up to 0.25). \ours~ achieves the highest review diversity among automated systems (RDS = 0.31), significantly reducing the gap to the natural variance of human committees (RDS = 0.43). Further, given the low variance in the win rates, these diverse reviews are of high quality. We hypothesize that \ours~ compounds diversity through its multi-agent design: multiple runs lead to slightly different, valid literature expansion paths and Q\&A probes, simulating the varied expertise of a diverse human reviewer panel.  

\subsection{Summary of Gains}
\begin{figure*}[t]
    \centering
    \tiny 
    
    \tcbset{
        arcstyle/.style={
            enhanced,
            colback=white,
            colframe=ARCBlue,
            colbacktitle=ARCBlue,
            coltitle=white,
            fonttitle=\bfseries\small,
            boxrule=0.6pt,
            arc=2mm,
            drop shadow=black!30!white,
            left=2mm, right=2mm, top=1mm, bottom=1mm,
            toptitle=1mm, bottomtitle=1mm,
            boxsep=1mm
        }
    }

    \begin{tcbitemize}[raster columns=1,      
                       raster row skip=3mm,   
                       arcstyle]              
        
        \tcbitem[title={Vs. DeepReviewer-14B (Fine-tuned Baseline)}]
            \begin{itemize}[leftmargin=*, noitemsep, label={}, topsep=0pt]
                \item \textcolor{green!60!black}{\faCheckCircle} \textbf{Advantage: SOTA Verification \& Integrity} \\
                \ours~ consistently refutes novelty claims by citing specific prior work (e.g., Lightman et al., TaskPrompter) that DeepReviewer ignores. It also detects fatal flaws like data leakage (training on validation sets) and circular evaluation protocols.
                
                \item \textcolor{green!60!black}{\faCheckCircle} \textbf{Advantage: Factual Precision} \\
                \ours~ avoids ``hallucinated critiques.'' While DeepReviewer frequently claims parameters are missing when they are explicitly stated, \ours~ correctly locates them and focuses on substantive inconsistencies.
                
            \end{itemize}
        
        \tcbitem[title={Vs. AI Scientist v2 (Agentic Baseline)}]
            \begin{itemize}[leftmargin=*, noitemsep, label={}, topsep=0pt]
                \item \textcolor{green!60!black}{\faCheckCircle} \textbf{Advantage: Theoretical Contextualization} \\
                \ours~ demonstrates deeper domain mastery by connecting papers to broader phenomena (e.g., ``Gradient Masking'' in adversarial defense or ``Grokking''), whereas AI Scientist often remains surface-level.
                
                \item \textcolor{green!60!black}{\faCheckCircle} \textbf{Advantage: Actionable Constructiveness} \\
                Instead of generic advice (``add complexity analysis''), \ours~ prescribes exact remedies, naming specific missing datasets (e.g., SR25, GSO) or control experiments required for validation.
                
                \item \textcolor{red!70!black}{\faTimesCircle} \textbf{Disadvantage: Internal Sanity Checking} \\
                AI Scientist v2 is occasionally better at catching internal inconsistencies (e.g., mathematically impossible probability ratios) because \ours~ prioritizes external verification over internal logic checks.
            \end{itemize}
            
    \end{tcbitemize}
    \caption{Qualitative analysis of \ours~ vs. top baselines. We summarize the key comparative advantages and disadvantages derived from the reasoning traces of the LLM Judge. \ours~ dominates in external verification (SOTA checking) and contextual depth, while AI Scientist v2 remains competitive on internal consistency checks.}
    \label{fig:qualitative_summary}
    \vspace{-7mm}
\end{figure*}

We aggregated the reasoning traces from our LLM judge to distill the systematic advantages and disadvantages of \ours~ compared to key baselines. We show the results in Figure~\ref{fig:qualitative_summary}. We provide review examples and summary of advantages against other baselines in Appendix~\ref{app:qualitative_examples} and the summarization prompt in Appendix~\ref{app:eval_prompts}.



\subsection{Ablation, Hyperparameter Sensitivity \& Deployment Practicality}
\label{sec:ablation}

\begin{figure}[ht]
    \centering
    \begin{minipage}[c]{0.48\textwidth}
        \centering
        \captionof{table}{Component ablations (SxS win rate against AI Scientist v2 (Gemini 3.1 Pro) on 200 ScholarEval papers).}
        \label{tab:ablation}
        \small 
        \begin{tabular}{lc}
        \hline
        \textbf{Configuration} & \textbf{Win Rate (\%)} \\
        \hline
        \textbf{Full \ours~} & \textbf{85} \\
        w/o Literature Review & 79 ($\downarrow$ 6\%) \\
        w/o Historian & 81 ($\downarrow$ 4\%) \\
        w/o Baseline Scout & 76 ($\downarrow$ 9\%) \\
        w/o Q\&A & 59 ($\downarrow$ 26\%) \\
        w/o Summary & 75  ($\downarrow$ 10\%) \\
        \hline
        \end{tabular}
    \end{minipage}\hfill
    \begin{minipage}[c]{0.48\textwidth}
        \centering
        \includegraphics[width=\linewidth]{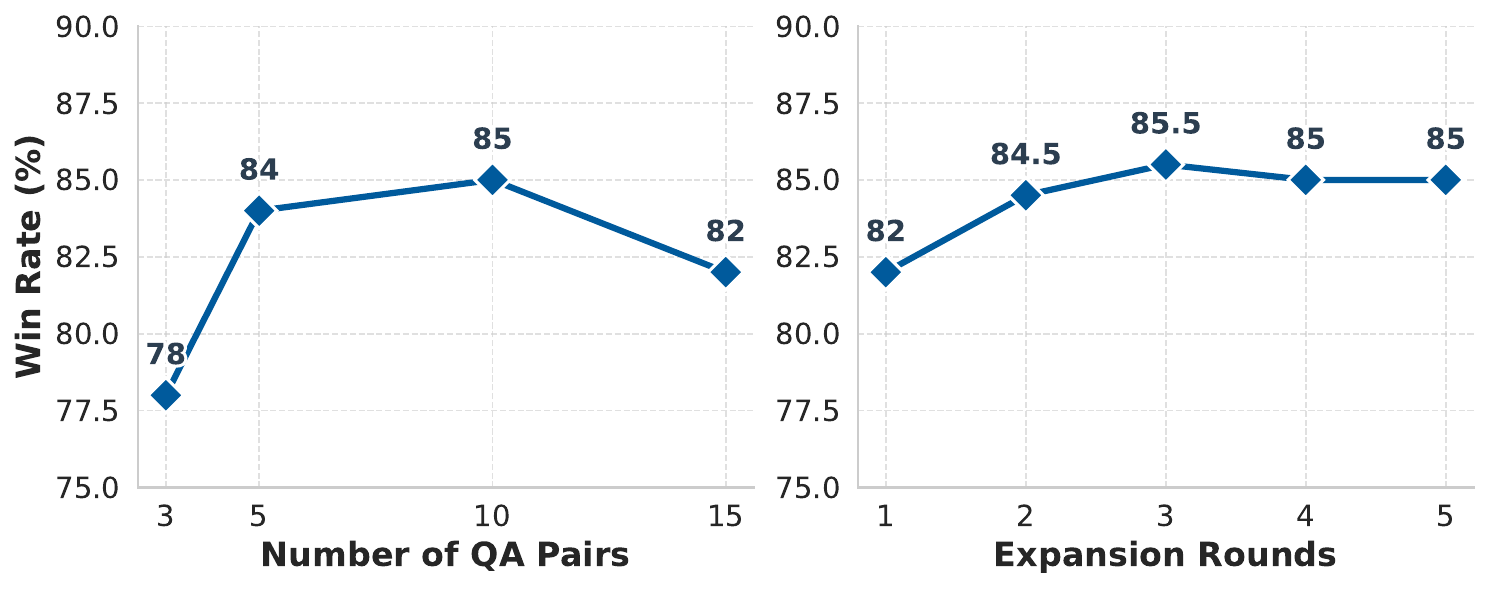}
        \captionof{figure}{Impact of search rounds ($k$) on review quality. Diminishing returns occur after $k=3$ as tangential papers dilute the context window.}
        \label{fig:hyperparam}
    \end{minipage}
    \vspace{-3mm}
\end{figure}

\textbf{Component Ablation.} We quantify the contribution of each module in Table \ref{tab:ablation}. Removing the baseline scout agent causes a 9\% drop in win-rate, as the system fails to identify missing comparisons. Removing the Q\&A agent causes the largest drop (26\%), crippling the system's ability to perform deep verification. The performance drop removing summarizer agent is surprisingly significant (10\%) signifying that internal compression also plays a vital role in reviewing.

\textbf{Hyperparameters.} We investigate the sensitivity of the system to the number of Q\&A pairs generated and the number of literature expansion rounds. As shown in Figure \ref{fig:hyperparam}, performance gains saturate after 10 Q\&A pairs and 3 search rounds, suggesting an optimal operating point.

\textbf{Deployment Practicality.} A critical consideration for multi-agent systems is operational overhead. \ours~ requires approximately 20 LLM calls per review, processing $\sim$429K input and $\sim$26K output tokens to actively retrieve and verify literature (see Appendix~\ref{app:complexity}). Using current enterprise API pricing, this equates to a monetary cost of $\sim$\$1.20 per paper. Single paper review latency is $\sim$10 min (reduced to amortized latency of $\sim$56 seconds per review with batch size 20). At the scale of major conferences (e.g., 10,000 submissions), this computational investment is highly justifiable given the immense reduction in manual verification hours for Area Chairs and reviewers.

\section{Related Work}

\textbf{LLM-based Automated Reviewing.} The field has rapidly evolved from text classification to complex generation tasks \citep{luo2025llm4sr}. Early approaches utilized standard pre-trained models for decision prediction and summarization \citep{liu2023reviewergpt, liang2024can}. Subsequent works shifted to supervised fine-tuning (SFT) on peer review datasets. \citet{weng2024cycleresearcher} introduced \textit{CycleReviewer} using negative constraints, while \textit{OpenReviewer} \citep{idahl2025openreviewer} and \textit{ReviewAgents} \citep{gao2025reviewagents} fine-tuned models to capture venue-specific stylistic norms. Most relevant to our work is \textit{DeepReviewer} \citep{zhu2025deepreview}, which employs chain-of-thought fine-tuning to improve reasoning.

\textbf{Agentic \& Retrieval-Augmented Frameworks.} To overcome static limitations, the field has pivoted toward agentic workflows \citep{wu2023autogen, bran2023chemcrow}. \citet{jin2024agentreview} pioneered this with \textit{Agent Review}, modeling the review process as a multi-stage discussion. Building on this collaborative approach, \textit{AgentRxiv} \citep{schmidgall2025agentrxiv} proposed a framework for autonomous research that facilitates interaction between various agents and human researchers throughout the scientific lifecycle. Other recent agentic reviewers such as \textit{ReviewRL} \citep{zeng2025reviewrl} and \textit{REMOR} \citep{taechoyotin2025remor} integrate reinforcement learning to align reviews with human preferences; REMOR utilizes GRPO with a multi-aspect reward function. \textit{AutoRev} \citep{chitale2025autorev} models papers as hierarchical graphs to optimize internal information retrieval, while \textit{PaperQA2} \citep{skarlinski2025paperqa2} uses agentic RAG on scientific Q\&A.


\textbf{Foundations of Automated Survey \& Verification.} Our architecture is grounded in the emerging literature of automated science. The historian agent builds upon automated survey generation frameworks, such as \textit{SurveyG} \citep{nguyen2025surveyg} and \textit{AutoSurvey} \citep{wang2024autosurvey}, which construct hierarchical citation graphs using LLMs. Similarly, the baseline scout operationalizes the ``chain-of-verification'' pattern \citep{dhuliawala2023cove}. This aligns with scientific fact-checking benchmarks like \textit{SciFact} and \textit{Co-Sight} \citep{zhang2025cosight}, moving beyond passive retrieval to actively identify omitted evidence and missing SOTA baselines.

\textbf{Evaluation of AI Reviewers.} Evaluating automated critiques remains challenging. \citet{garg2025revieweval} proposed \textit{ReviewEval} to assess actionability, and \textit{MMReview} \citep{gao2025mmreview} introduced a multimodal benchmark. Standard metrics, however, often fail to capture systemic automation risks. \textit{LLM-REVal} \citep{li2025llmreval} highlighted LLM judge bias, while \citet{jiang2025artificial} identified the risk of an ``artificial hivemind'' homogenizing opinions. Our work complements these efforts by prioritizing human-calibrated actionability via the \hmaxmetric~score and explicitly measuring homogenization risks through the review diversity score.
\section{Conclusion}
In this paper, we introduced \ours, a multi-agent framework that structurally decouples intrinsic paper comprehension from external contextualization, and operationalizes the rigorous workflow of a senior researcher. Our specialized agents dynamically synthesize domain trajectories, proactively hunt for omitted baselines, and deeply interrogate technical soundness. Crucially, \ours~ is designed not to replace human judgment, but to serve as a ``co-scientist'' that accelerates author iteration and augments human reviewers with highly actionable, verifiable evidence. 

Our extensive evaluation on $\sim$1,800 ICLR submissions spanning 2020 through 2025 demonstrates that \ours~ sets a new standard for automated critique. By achieving an \hmaxmetric~score of 6.12 and decisively outperforming both state-of-the-art fine-tuned models and search-augmented agentic baselines, we empirically prove that structured adversarial reasoning—rather than mere internet access—is the essential catalyst for expert-level review. Future work will explore collaborative multi-agent debate mechanisms to further deepen the rigor, diversity, and actionability of artificial scientific critique.
\newpage

\bibliographystyle{abbrvnat}
\nobibliography*
\bibliography{main}

\clearpage

\appendix

\newpage
\appendix
\onecolumn

\section{Impact Statement}
This paper introduces \ours, an active verification framework for automated peer review—a domain with significant potential for both positive and negative societal impact. On the positive side, our work addresses the critical scalability crisis in scientific publishing. By operationalizing the rigorous stress-testing workflow of a senior researcher, \ours~ moves beyond passive text generation to provide deep, verifiable critiques. This aims to democratize access to high-quality, actionable feedback, accelerating scientific progress particularly for authors who may lack access to senior mentorship.

However, we acknowledge several ethical risks associated with the deployment of automated evaluation systems:

\begin{itemize}
    \item \textbf{Automation Bias:} There is a risk that human area chairs or reviewers may over-rely on AI-generated critiques, accepting them as authoritative without independent scrutiny. While \ours's design explicitly anchors claims in verifiable literature through its interrogation logs—avoiding the hallucinations of passive summarization—the risk of human complacency remains a critical deployment challenge.
    \item \textbf{Homogenization of Science:} Automated systems might penalize novel ideas that do not conform to established patterns in their training data or the retrieved literature. Although we explicitly designed the multi-agent architecture to maximize perspective variance and introduced the \textit{Review Diversity Score} to monitor this issue, widespread reliance on automated models could inadvertently narrow the scope of acceptable scientific inquiry.
    \item \textbf{Privacy and Confidentiality:} The use of search-enabled agents involves processing unpublished manuscripts via external LLM and search APIs. Any real-world deployment must strictly adhere to venue confidentiality policies, ensuring that submission data is secure, ephemeral, and never used to train public models.
\end{itemize}

We emphasize that \ours~ is designed strictly to augment, not replace, human judgment. By acting as a ``co-scientist''---a rigorous mentor for authors and a tireless verification assistant for reviewers---the system's output should be treated as an evidence-grounded second opinion, ensuring that final scientific decisions remain firmly in human hands.

\section{Limitations}
\label{app:limitations}

While ScholarPeer represents a significant advancement, several limitations remain. First, despite improving upon baselines, the review diversity of our system (0.31) still lags behind the high variance of human expert reviews (0.43); closing this gap is crucial for capturing the full spectrum of scientific opinion. Second, the reliance on active web-scale retrieval introduces higher inference latency and cost compared to static fine-tuned models.

\section{Experiment Details}
\label{app:experiment_details}

To ensure reproducibility, we provide the specific configuration details for the \ours~ framework and baselines.

\subsection{Models and Compute}
All experiments involving \ours~ were conducted using the Google Cloud Vertex AI platform.
\begin{itemize}
    \item \textbf{Backbone Model:} We utilized \texttt{Gemini 3.1 Pro} as the core reasoning engine for all agents in the primary ScholarPeer configuration.
    \item \textbf{Temperature:} We set the temperature to \texttt{0.7} for all generation tasks to balance creativity with adherence to instructions.
    \item \textbf{Compute Infrastructure:} Experiments were orchestrated on standard CPU instances for agentic frameworks, with all heavy lifting offloaded to the Vertex AI Model-as-a-Service (MaaS) endpoints. We used 8X NVIDIA A100 80GB GPUs for fine-tuned models (CycleReviewer and DeepReview).
    \item \textbf{Search Tools:} For Gemini Models, we utilized the native \texttt{google\_search} tool provided by the Vertex AI Gemini API. This allows the model to perform multi-step reasoning and retrieve up-to-date information from the open web. For Claude Models, we employed the \texttt{web\_search\_20250305} tool to ensure a fair comparison with a similarly capable search-enabled environment.
\end{itemize}

\subsection{Inter-Judge Agreement and Correlation}
\begin{table}[ht]
\centering
\caption{Pairwise agreement between LLM judges on the SxS evaluation (overall dimension).}
\label{tab:judge_agreement}
\begin{tabular}{lc}
\toprule
Judge Pair & Agreement Rate \\
\midrule
Claude Sonnet 4.5 vs Gemini 3.1 Flash & 0.79 \\
Claude Sonnet 4.5 vs Gemini 3.1 Pro & 0.77 \\
Gemini 3.1 Flash vs Gemini 3.1 Pro & 0.88 \\
\bottomrule
\end{tabular}
\end{table}

Table \ref{tab:judge_agreement} provides the pairwise agreement between LLM judges on the SxS evaluation (overall dimension). Gemini 3.1 Flash and 3.1 Pro have the highest agreement rate. This is intuitive as they are from the same model family. Claude Sonnet 4.5 and Gemini 3.1 Pro have the lowest agreement rate but the agreement is still high (77\%).

\begin{table}[ht]
\centering
\caption{Average correlation between the scores provided by the LLM judges for \hmaxmetric~score.}
\label{tab:judge_correlation}
\begin{tabular}{lc}
\toprule
Judge Pair & \hmaxmetric~Score Correlation \\
\midrule
Claude Sonnet 4.5 vs Gemini 3.1 Flash & 0.62 \\
Claude Sonnet 4.5 vs Gemini 3.1 Pro & 0.66 \\
Gemini 3.1 Flash vs Gemini 3.1 Pro & 0.81 \\
\bottomrule
\end{tabular}
\end{table}

Table \ref{tab:judge_correlation} provides the average correlation between the \hmaxmetric~scores provided by the LLM judges. Similar to agreement rate, Gemini 3.1 Flash and Pro have the highest score correlation and Claude Sonnet 4.5 has lower correlation with Gemini models but the correlation is high (0.66 with Gemini 3.1 Pro).

\section{Computational Complexity and Deployment Analysis}
\label{app:complexity}

In this section, we analyze the computational overhead of the \ours~ framework, covering theoretical LLM inference calls and empirical token usage, monetary cost, and deployment latency. We demonstrate that \ours~ is highly scalable and practical for conference-level deployment.

\subsection{Theoretical Inference Cost}
The computational cost of \ours~ is driven by its multi-agent architecture, specifically the depth of the literature search and the rigor of the Q\&A verification. The total number of inference calls is defined by a fixed overhead for context construction plus variable costs associated with literature expansion rounds ($k$) and the number of probing questions generated ($N_{QA}$):

\begin{equation}
    C_{\text{ScholarPeer}} \approx C_{\text{fixed}} + k + N_{QA}
\end{equation}

The fixed overhead ($C_{\text{fixed}} \approx 7$) comprises single calls for the summary agent, initial literature search, historian, baseline scout, aspect-based question generation (novelty and soundness), and the final review generator. Based on our hyperparameter sensitivity analysis, we utilize $k=3$ expansion rounds and generate $N_{QA}=10$ probing questions to maximize performance while containing costs. This configuration results in approximately \textbf{20 LLM calls} per paper review, compared to a single forward pass ($C=1$) for static fine-tuned models (e.g., DeepReviewer) or 5--10 calls for standard agentic baselines (e.g., AI Scientist v2).

\subsection{Empirical Token Usage and Monetary Cost}
To calculate the practical deployment cost of \ours, we aggregated the total token consumption across all internal agent calls over the ScholarEval test set. Because the system actively retrieves and compresses multiple external literature sources, an average \ours~ review consumes approximately $\sim$429K input tokens and $\sim$26K output tokens per manuscript.

Using standard enterprise API pricing for our primary backbone model, Gemini 3.1 Pro (\$2.00 per 1M input tokens and \$12.00 per 1M output tokens)\footnote{\url{https://cloud.google.com/vertex-ai/generative-ai/pricing}}, this equates to a monetary cost of \textbf{$\sim$\$1.20 per review}.

\subsection{Latency and Scalability}
We evaluated the system's runtime latency to ensure viability for major machine learning conferences, which frequently receive over 10,000 submissions annually.
\begin{itemize}
    \item \textbf{Single-Paper Latency:} Without batching, executing the full parallelized multi-agent workflow—including web retrieval, internal extraction, and active interrogation—for a single manuscript takes approximately \textbf{$\sim$10 minutes}.
    \item \textbf{Amortized Latency:} When processing papers in parallel batches (using 20 concurrent threads), a full run on the DeepReview-13K test set (1,286 papers) completes in $\sim$20 hours. This yields a highly scalable amortized latency of just \textbf{$\sim$56 seconds per paper}.
\end{itemize}

\textbf{Summary:} While \ours~ requires more computational resources than static single-pass models, those models fundamentally lack the structural capability to perform external literature searches or actively verify claims. The $\sim$\$1.20 cost and sub-minute amortized latency represent a negligible operational overhead when compared to the hundreds of human hours required to manually audit missing baselines and verify technical soundness, making \ours~ an exceptionally practical augmentation tool for real-world academic deployment.

\section{Qualitative Examples}
\label{app:qualitative_examples}

We present examples of reviews generated by baselines and compare them to \ours~. These examples highlight the system's ability to identify missing baselines and verify novelty.

\subsection{Example 1: Detection of Missing Baselines}
\label{app:example1}
\textbf{Paper Title:} DBRNet: Advancing Individual-Level Continuous Treatment Estimation through Disentangled and Balanced Representation \\
\textbf{OpenReview Link:} \url{https://openreview.net/pdf?id=tlqmkftgpw}

\begin{tcolorbox}[colback=gray!5, colframe=gray!40, title=Baseline Review (DeepReviewer-14B), breakable]
\footnotesize
\textbf{Summary:}\newline
This paper introduces DBRNet, a novel neural network architecture designed for estimating individualized dose-response functions (IDRFs) in the presence of continuous treatments. The core contribution of DBRNet lies in its integration of three key components: disentangled representation learning, balancing via re-weighting, and a varying coefficient network. The method begins by learning disentangled representations of covariates, aiming to separate instrumental, confounding, and adjustment factors. This disentanglement is achieved through a combination of a discrepancy loss and an independent loss. To address selection bias, DBRNet employs a re-weighting function based on the learned representations, which aims to create a more balanced comparison group. Finally, a varying coefficient network is used to model the continuous treatment effect, allowing for flexible modeling of the dose-response relationship. The authors provide theoretical justification for their debiasing approach, arguing that the combination of disentangled representations and re-weighting effectively mitigates selection bias. Empirically, the authors demonstrate the effectiveness of DBRNet on both synthetic and semi-synthetic datasets, showing that it outperforms several state-of-the-art methods in terms of mean integrated squared error (MISE) and average mean squared error (AMSE). The paper presents a comprehensive framework for IDRF estimation, combining several existing ideas in a novel way. While the empirical results are promising, the paper also has several limitations that warrant further discussion. The paper's focus on methodological innovation is commendable, but a more thorough discussion of practical considerations and limitations would enhance its overall impact. The paper's contribution is significant in its attempt to address the challenges of continuous treatment effects, but the identified weaknesses need to be addressed to ensure the robustness and generalizability of the proposed method. The paper's approach to disentangled representation learning and re-weighting is a valuable contribution to the field, but the lack of discussion on computational complexity, missing data, and high-dimensional covariates limits its practical applicability. The paper's theoretical justification is a strength, but the empirical evaluation could be strengthened by including more diverse datasets and a more detailed analysis of the method's sensitivity to various data characteristics. Overall, the paper presents a valuable contribution to the field of individualized treatment effect estimation, but it also highlights several areas where further research is needed.
\par\vspace{0.5em}

\textbf{Strengths:}\newline
The paper presents a well-structured and clearly articulated approach to estimating individualized dose-response functions (IDRFs) with continuous treatments. The core strength of the paper lies in its novel integration of disentangled representation learning, balancing via re-weighting, and a varying coefficient network. This combination, while drawing inspiration from existing methods, is a unique contribution that addresses the challenges of continuous treatment effects in a comprehensive manner. The theoretical justification for the debiasing approach is another significant strength. The authors provide a clear argument for how the proposed method mitigates selection bias through the use of disentangled representations and re-weighting, which adds a layer of rigor to the proposed methodology. The empirical results, while limited in scope, demonstrate the effectiveness of DBRNet on both synthetic and semi-synthetic datasets. The paper shows that DBRNet achieves superior performance compared to several state-of-the-art methods in terms of MISE and AMSE, which provides empirical support for the proposed approach. The ablation study further strengthens the empirical findings by demonstrating the importance of the re-weighting function and the discrepancy loss, highlighting the contribution of the specific combination of components. The paper's focus on addressing the challenges of continuous treatments is also a strength, as this is an important and often overlooked aspect of individualized treatment effect estimation. The paper's clear presentation of the method, including the detailed description of the objective function and its components, makes it easy to follow and understand. The use of a varying coefficient network to model the continuous treatment effect is a technically sound approach that allows for flexible modeling of the dose-response relationship. The paper's overall contribution is significant in that it provides a comprehensive framework for IDRF estimation that combines several existing ideas in a novel and effective way. The paper's focus on addressing selection bias through disentangled representations and re-weighting is a valuable contribution to the field, and the empirical results provide initial evidence of the method's effectiveness.
\par\vspace{0.5em}

\textbf{Weaknesses:}\newline
Upon careful examination, several weaknesses in the paper warrant attention. First, the paper lacks a discussion on the generalizability of DBRNet to datasets with missing covariates. The method description and experimental setup assume complete data, and there is no mention of how missing data would be handled in the disentanglement process or the re-weighting function. This is a significant limitation, as missing data is a common issue in real-world observational studies. The paper does not address how the disentanglement or re-weighting would be affected by missing data, especially under different missing data mechanisms. This omission raises concerns about the practical applicability of DBRNet in real-world scenarios where missing data is a common issue. My confidence in this weakness is high, as the paper's method description and experimental setup explicitly assume complete data, and there is no discussion of handling missing values. Second, the paper does not provide a detailed analysis of the time complexity of DBRNet. The paper describes the components of DBRNet but does not provide a formal analysis of their computational cost. The experimental section lacks any timing information or comparisons of training/inference time with other methods. This is a significant oversight, as the computational cost of a method is a crucial factor in its practical applicability, especially on large-scale datasets. Without a time complexity analysis, it is difficult to assess the feasibility of applying DBRNet to real-world problems. My confidence in this weakness is high, as the paper lacks any discussion or empirical evaluation of the computational cost of DBRNet. Third, the paper also lacks a discussion of the space complexity of DBRNet. The paper describes the network architecture but does not provide a formal analysis of the memory requirements for storing the model parameters and intermediate representations. The experimental section does not provide any information about memory usage. This is a significant limitation, as the memory footprint of a method is a crucial factor in its practical applicability, especially in resource-constrained environments. Without a space complexity analysis, it is difficult to assess the feasibility of applying DBRNet in resource-limited settings. My confidence in this weakness is high, as the paper lacks any discussion or empirical evaluation of the memory requirements of DBRNet. Fourth, the paper does not adequately address the limitations of DBRNet when dealing with high-dimensional covariates. While the paper uses neural networks, which can be prone to overfitting in high-dimensional settings, there is no specific discussion about the challenges of high-dimensional data. The datasets used in the experiments have a moderate number of features, and the paper does not include experiments on datasets with a very large number of covariates. This is a significant limitation, as high-dimensional data is common in many real-world applications. The paper does not discuss potential strategies for mitigating the curse of dimensionality, such as dimensionality reduction techniques or regularization methods. My confidence in this weakness is high, as the paper lacks any discussion of the performance and potential limitations of DBRNet on high-dimensional datasets. Finally, while the paper touches upon relevant concepts like selection bias, it lacks an explicit and detailed discussion of methods tailored for observational studies. While the paper compares against methods that can be used in observational settings, it does not explicitly frame the discussion around the specific challenges and solutions prevalent in the observational causal inference literature. This omission limits the paper's ability to contextualize its contribution within the broader field of causal inference. My confidence in this weakness is medium, as the paper does discuss methods relevant to observational settings, but it lacks an explicit and detailed discussion of the specific literature on ITE estimation in observational studies. Additionally, the paper does not provide a detailed comparison of DBRNet with transformer-based and GAN-based DRF estimation methods, particularly regarding long-range dependencies and mode collapse. This lack of comparison limits the paper's ability to contextualize its contribution within the broader field of causal inference. My confidence in this weakness is high, as the related work section lacks a detailed comparison of DBRNet's architecture and assumptions with transformer-based and GAN-based DRR estimation methods. Furthermore, it is unclear whether DBRNet uses targeted regularization, and if so, how it compares to the baselines with targeted regularization. Providing results without targeted regularization for DBRNet would ensure a fairer comparison. My confidence in this weakness is high, as the paper does not explicitly state whether DBRNet uses targeted regularization, making it unclear how the comparison with baselines using targeted regularization is fair. Results without targeted regularization for DBRNet are missing.
\par\vspace{0.5em}

\textbf{Suggestions:}\newline
To address the identified weaknesses, I recommend several concrete improvements. First, the authors should thoroughly investigate the performance of DBRNet on datasets with missing covariates. This should involve conducting experiments on datasets with simulated missing values under different assumptions (missing completely at random, missing at random, missing not at random) and analyzing how the disentanglement and re-weighting processes are affected. The authors should also explore potential strategies for handling missing data, such as imputation techniques, and how these strategies might interact with DBRNet's performance. This analysis should include both theoretical considerations and empirical evaluations on datasets with varying degrees and patterns of missingness. Second, the authors should provide a detailed analysis of the time complexity of DBRNet. This should include an analysis of the computational cost associated with each step of the algorithm, such as the disentanglement of representations, the calculation of the re-weighting function, and the outcome prediction. A comparison with the computational complexity of other state-of-the-art methods would be beneficial for assessing the practical applicability of DBRNet, especially on large-scale datasets. Additionally, the authors should discuss the practical implications of the computational complexity, such as the feasibility of applying DBRNet to large-scale datasets and the potential need for parallelization or other optimization techniques. Third, the authors should provide a detailed analysis of the space complexity of DBRNet. This should include an analysis of the memory requirements for storing the model parameters and intermediate representations. The authors should also discuss how the memory footprint scales with the size of the input data and the complexity of the model. This analysis should consider the impact of different hyperparameter settings on the memory requirements. Fourth, the authors should address the limitations of DBRNet when dealing with high-dimensional covariates. This should include a discussion of the potential challenges associated with high-dimensional data, such as the curse of dimensionality and the increased risk of overfitting. The authors should explore potential solutions, such as dimensionality reduction techniques or regularization methods, to mitigate these challenges. Furthermore, the authors should discuss the impact of high dimensionality on the disentanglement process and the accuracy of the treatment effect estimation. It would be beneficial to include experiments on datasets with varying numbers of covariates to demonstrate the scalability of DBRNet and to identify any potential limitations. Fifth, the authors should expand the related work section to include a more comprehensive discussion of methods specifically designed for estimating individualized continuous treatment effects in observational studies. This should include a discussion of the specific challenges and solutions prevalent in the observational causal inference literature. The authors should also provide a more detailed comparison of DBRNet with transformer-based and GAN-based DRF estimation methods, particularly regarding long-range dependencies and mode collapse. This should include a discussion of the specific architectural differences, such as the use of attention mechanisms in transformers and how they handle long-range dependencies in the covariate space, compared to the proposed method. It should also address the challenges of training GANs for continuous treatment settings, such as mode collapse, and how the proposed method avoids these issues. Finally, the authors should clarify whether DBRNet uses targeted regularization and, if so, which loss terms are targeted. If DBRNet does not use targeted regularization, the authors should provide results without targeted regularization to ensure a fair comparison with baselines that use targeted regularization. These changes would significantly improve the paper's clarity, rigor, and practical relevance.
\par\vspace{0.5em}

\textbf{Questions:}\newline
Several key questions arise from my analysis of the paper. First, how does the performance of DBRNet degrade when applied to datasets with varying degrees of missingness, and what are the best strategies for handling missing data in the context of the proposed method? Specifically, how do different imputation techniques interact with the disentanglement and re-weighting processes? Second, what is the precise time complexity of each component of DBRNet, and how does it compare to the time complexity of other state-of-the-art methods for IDRF estimation? What are the practical implications of this computational cost, especially when applied to large-scale datasets? Third, what is the space complexity of DBRNet, and how does the memory footprint scale with the size of the input data and the complexity of the model? Are there any specific optimizations that can be implemented to reduce the memory requirements of DBRNet? Fourth, how does the performance of DBRNet change when applied to datasets with high-dimensional covariates, and what are the best strategies for mitigating the curse of dimensionality in this context? Are there specific dimensionality reduction techniques or regularization methods that are particularly effective for DBRNet? Fifth, how does DBRNet compare to other methods specifically designed for observational studies, and what are the key differences in their assumptions and approaches? How does DBRNet address the specific challenges of confounding and selection bias in observational settings? Sixth, how does DBRNet's architecture compare to transformer-based and GAN-based DRF estimation methods in terms of handling long-range dependencies and mode collapse? What are the specific advantages and disadvantages of each approach? Finally, does DBRNet use targeted regularization, and if so, which loss terms are targeted? If not, how does the performance of DBRNet compare to the baselines when the baselines are not using targeted regularization? These questions highlight key uncertainties and areas where further clarification and investigation are needed to fully understand the strengths and limitations of the proposed method.
\end{tcolorbox}

\begin{tcolorbox}[colback=white, colframe=black, title=\ours Review , breakable]
\footnotesize
\textbf{Summary:}\newline
The paper proposes DBRNet (Disentangled and Balanced Representation Network) for estimating Individualized Dose-Response Functions (IDRF) in continuous treatment settings. The method addresses selection bias by disentangling covariates into three latent factors: Instrumental, Confounder, and Adjustment factors. It employs three specific neural networks to extract these factors, using only the Instrumental and Confounder factors to estimate propensity scores for re-weighting, and the Confounder and Adjustment factors to predict outcomes via a Varying Coefficient Network. The authors introduce a specific loss function to enforce independence between adjustment factors and treatment, and provide theoretical proofs for the unbiasedness of their re-weighting scheme.
\par\vspace{0.5em}

\textbf{Strengths:}
\begin{itemize}
    \item \textbf{Theoretical Motivation}: The core idea of 'selective balancing' specifically avoiding the balancing of instrumental and adjustment factors is well-grounded in causal inference theory and addresses a known limitation of methods that balance the entire covariate representation.
    \item \textbf{Ablation Studies}: The ablation study effectively demonstrates the contribution of individual components, particularly the discrepancy loss and re-weighting module, to the model's performance on the chosen datasets.
\end{itemize}
\par\vspace{0.5em}

\textbf{Weaknesses:}
\begin{itemize}
    \item \textbf{Missing Critical Baselines}: The paper claims that no existing efforts are capable of precisely adjusting for selection bias in continuous settings. This overlooks significant recent contributions such as \textbf{ADMIT (NeurIPS 2022)} and \textbf{End-to-End Balancing (ICML 2022)}, which explicitly focus on re-weighting and balancing for continuous treatments. The comparison is largely limited to methods from 2021 (VCNet) and earlier.
    \item \textbf{Missing Standard Benchmark}: The evaluation excludes the \textbf{TCGA} dataset, which is a standard high-dimensional semi-synthetic benchmark for continuous treatment effect estimation used by VCNet, SCIGAN, TransTEE, and ADMIT. The omission of this challenging high-dimensional dataset limits the assessment of the method's robustness.
    \item \textbf{Problematic Independence Loss ($L_{ind}$)}: The proposed mechanism to enforce independence (minimizing the log-likelihood of treatment given adjustment factors) is mathematically brittle. Maximizing prediction error is unbounded and does not guarantee that the representation contains no information about the treatment (mutual information minimization), only that it predicts the specific observed treatment poorly. Standard approaches utilize Adversarial Training (Gradient Reversal Layers) or mutual information minimization to ensure true independence.
    \item \textbf{Naive Density Estimation}: Despite focusing on continuous treatments, the propensity score estimation relies on discretizing the treatment space into grids and using linear interpolation. This introduces approximation errors and seems inconsistent with the goal of handling continuous treatments rigorously (e.g., compared to Mixture Density Networks or Normalizing Flows).
    \item \textbf{Lack of Balance Metrics}: The paper evaluates performance solely on outcome prediction error (MISE/AMSE). It lacks metrics evaluating covariate balance (e.g., weighted correlations) or the accuracy of the propensity estimation itself, which are crucial for verifying the claim of 'precise bias adjustment'.
\end{itemize}
\par\vspace{0.5em}

\textbf{Suggestions:}
\begin{itemize}
    \item Include comparisons with ADMIT (NeurIPS 2022) and End-to-End Balancing (ICML 2022) to substantiate claims of state-of-the-art performance.
    \item Evaluate the method on the TCGA dataset to demonstrate robustness in high-dimensional settings.
    \item Replace the $L_{ind}$ maximization of error with a more stable independence constraint, such as a Gradient Reversal Layer (GRL) or minimizing Mutual Information.
    \item Implement a truly continuous density estimator (e.g., Normalizing Flows or MDNs) for the propensity score to align with the continuous treatment setting.
    \item Report metrics on covariate balance (e.g., weighted correlation between confounders and treatment) to empirically verify the 'precise adjustment' claim.
\end{itemize}
\par\vspace{0.5em}

\textbf{Questions:}
\begin{itemize}
    \item Why were recent state-of-the-art re-weighting methods for continuous treatments, specifically ADMIT (NeurIPS 2022) and End-to-End Balancing (ICML 2022), omitted from the baselines?
    \item Why was the standard TCGA benchmark excluded from the evaluation, given its prevalence in the literature (used by VCNet, TransTEE, etc.)?
    \item Minimizing the log-likelihood (maximizing negative log-likelihood) is unbounded. How do you prevent the encoder weights from exploding or the distribution from shifting to infinity to satisfy this loss? Why was this chosen over standard Adversarial Training (Gradient Reversal Layer)?
    \item The treatment density estimator uses discretization (grids). How sensitive is the performance to the number of grids ($B$), and why was a continuous density estimator (like MDN) not used?
\end{itemize}
\par\vspace{0.5em}
\end{tcolorbox}

\begin{tcolorbox}[colback=blue!5!white, colframe=blue!50!black, title=LLM-as-a-Judge Evaluation (Gemini 3.1 Pro), breakable]
\footnotesize
\textbf{Winner:} \ours~ 

\textbf{Technical Accuracy Reason:}
ScholarPeer correctly identifies that the paper ignores critical recent SOTA methods (\textit{ADMIT}, \textit{End-to-End Balancing}) and standard benchmarks (\textit{TCGA}) that contradict the authors' claims. The Baseline focuses on generic issues like missing data and complexity which are less central to the paper's scientific validity.

\textbf{Analytical Depth Reason:}
ScholarPeer critiques the mathematical formulation of the loss function (unbounded maximization of error) and the inconsistency of using discrete grid-based density estimation for a continuous treatment method. DeepReview 14B does not engage with the technical details of the architecture or loss functions.

\textbf{Novelty \& Significance Reason:}
ScholarPeer correctly identifies that the paper's claim of being the 'first' to precisely adjust for selection bias in continuous settings is false, citing specific prior works (ADMIT, End-to-End Balancing) published in 2022. DeepReview 14B accepts the authors' claims without verifying the existence of recent competing methods.

The following external sources found during the search helped verify this:
\begin{itemize}[leftmargin=*, nosep]
    \item \textit{Bahadori, M. T., Tchetgen, E. T., \& Heckerman, D. (2022). End-to-End Balancing for Causal Continuous Treatment-Effect Estimation. ICML.}
    \item \textit{Wang, X., Lyu, S., Wu, X., Wu, T., \& Chen, H. (2022). Generalization Bounds for Estimating Causal Effects of Continuous Treatments. NeurIPS.}
\end{itemize}

\textbf{Constructive Value Reason:}
ScholarPeer provides specific, actionable steps to bring the paper up to scientific standards: compare against named baselines, use the TCGA dataset, and fix the unstable loss using Gradient Reversal Layers. The Baseline's suggestions (``analyze time complexity'') are generic and apply to almost any deep learning paper.

\textbf{Overall Verdict:}
ScholarPeer is superior in every dimension. It identifies a fatal flaw (ignoring direct prior work), points out the omission of a standard benchmark, and offers deep technical critiques of the loss function. The Baseline provided a superficial review with generic comments.
\end{tcolorbox}

\subsection{Example 2: Technical Soundness}
\textbf{Paper Title:} GAN-based Vertical Federated Learning for Label Protection \\
\textbf{OpenReview Link:} \url{https://openreview.net/pdf?id=5Vh0XqOTGi}

\begin{tcolorbox}[colback=gray!5, colframe=gray!40, title=Baseline Review (AI Scientist v2), breakable]
\footnotesize
\textbf{Summary:}\newline
The paper proposes the Generative Adversarial Federated Model (GAFM) to mitigate Label Leakage from Gradients (LLG) in Vertical Federated Learning (VFL) based on SplitNN. The method replaces the standard label-party architecture with a GAN, where a generator predicts labels and a discriminator aligns them with the true distribution. A randomized Cross-Entropy loss is added to maintain utility. The approach relies on the mutual perturbation of gradients from the GAN and CE objectives to obscure label information. The authors also introduce two new gradient-based attacks (Mean and Median attacks) and demonstrate that GAFM achieves a better privacy-utility trade-off than baselines like Marvell and Max Norm on binary classification tasks.
\par\vspace{0.5em}

\textbf{Strengths:}
\begin{itemize}
    \item \textbf{Novelty}The integration of a GAN-based architecture specifically for privacy preservation in SplitNN is a novel design choice.
    \item \textbf{Empirical Performance}: The method demonstrates a superior trade-off between model utility (AUC) and privacy (Leak AUC) compared to strong baselines like Marvell. It also shows better stability across random seeds.
    \item \textbf{Evaluation Metrics}: The proposal of Mean and Median attacks provides a stronger and more rigorous benchmark for evaluating gradient separability compared to the standard Norm attack.
    \item \textbf{Ablation Studies}: The necessity of both the GAN component (for distribution matching) and the randomized CE component (for instance-level accuracy) is well-supported by ablation studies.
\end{itemize}
\par\vspace{0.5em}

\textbf{Weaknesses:}
\begin{itemize}
    \item \textbf{Lack of Theoretical Guarantees}: The defense is heuristic. Unlike some baselines that offer bounds based on information theory (e.g., KL divergence), GAFM relies on empirical validation of gradient mixing.
    \item \textbf{Computational Overhead}: The paper does not sufficiently analyze the training time and computational resources required for the GAN components compared to standard SplitNN or simpler defenses.
    \item \textbf{Attack Assumptions}: The proposed attacks assume the attacker knows the gradient centers or medians of the classes. While useful for measuring separability, this may not fully reflect realistic scenarios where attackers must estimate these via clustering.
    \item \textbf{Hyperparameter Tuning}: The suggested strategy of tuning parameters on a data subset may be non-trivial to implement in a strict VFL setting without violating privacy constraints.
\end{itemize}
\par\vspace{0.5em}

\textbf{Limitations:}
\begin{itemize}
    \item The method lacks formal theoretical privacy bounds.
    \item Current evaluation is limited to binary classification.
    \item Computational complexity is higher than simple noise-addition methods.
\end{itemize}
\par\vspace{0.5em}

\textbf{Questions:}
\begin{itemize}
    \item What is the quantitative computational overhead (time per epoch, memory) of GAFM compared to Marvell and vanilla SplitNN?
    \item Regarding the randomized response mechanism: Since the target intervals for class 0 and 1 are disjoint (for $\\Delta < 0.5$), how does the method theoretically ensure that the GAN gradient sufficiently masks the CE gradient component to prevent label recovery?
    \item Can the authors clarify how the hyperparameter tuning on a subset can be performed securely in a real-world VFL deployment?
    \item Do the Mean/Median attacks perform similarly if the attacker estimates centers using unsupervised clustering (e.g., K-Means) instead of using ground truth statistics?
\end{itemize}
\par\vspace{0.5em}
\end{tcolorbox}

\begin{tcolorbox}[colback=white, colframe=black, title=\ours Review , breakable]
\footnotesize
\textbf{Summary:}\newline
The paper introduces the Generative Adversarial Federated Model (GAFM) to address Label Leakage from Gradients (LLG) in Vertical Federated Learning (specifically Split Learning). The authors propose replacing the standard server-side model with a GAN architecture, where a Generator acts as the predictor and a Discriminator aligns the prediction distribution with the label distribution. The training objective combines a Wasserstein GAN loss with a Randomized Cross-Entropy (CE) loss (using Randomized Response). The authors argue that this combination creates 'mutual gradient perturbation,' where the conflicting gradients from the GAN and CE components obfuscate the true label information in the gradients sent to the client. The paper also introduces two new gradient clustering attacks (Mean and Median) and evaluates the method on four binary classification datasets.
\par\vspace{0.5em}

\textbf{Strengths:}
\begin{itemize}
    \item \textbf{Problem Relevance:}  The paper addresses a critical privacy vulnerability (Label Leakage from Gradients) in Split Learning, which is a significant barrier to the adoption of Vertical Federated Learning.
    \item \textbf{New Attack Baselines:} The proposed Mean and Median attacks appear to be more effective than the standard Norm attack, providing a stronger empirical baseline for evaluating future defenses.
    \item \textbf{Empirical Performance:} On the tested binary classification datasets, GAFM demonstrates a favorable trade-off between model utility (AUC) and privacy (Leak AUC) compared to the selected baselines (Marvell, Max Norm).
\end{itemize}
\par\vspace{0.5em}

\textbf{Weaknesses:}
\begin{itemize}
    \item \textbf{Lack of Theoretical Guarantees:} A major weakness is the absence of formal privacy guarantees. Unlike Marvell, which provides a theoretical upper bound on leakage based on KL-divergence, or Differential Privacy methods, GAFM relies on a heuristic argument of 'mutual gradient perturbation.' In privacy research, empirical defense against specific attacks is often insufficient as it does not guarantee protection against adaptive or future attacks.
    \item \textbf{Limited Scope (Binary Only):} The proposed mechanism relies heavily on the geometry of binary classification, where gradients for positive and negative classes tend to pull in opposite directions. The paper explicitly acknowledges this limitation. This severely restricts the significance of the work, as the method does not appear generalizable to multi-class problems where gradient interactions are high-dimensional and complex.
    \item \textbf{Missing Baselines:} The evaluation is restricted to perturbation-based defenses. The paper fails to compare against state-of-the-art non-perturbation defenses such as Confusional AutoEncoders (CoAE) or deterministic gradient quantization methods (e.g., DiscreteSGD-like approaches), which are relevant to the LLG problem.
    \item \textbf{Questionable Stability Claims:} The paper claims GAFM is more stable than Marvell. However, GANs are notoriously difficult to train and prone to instability (mode collapse, vanishing gradients). The defense relies on a hyperparameter $\\gamma$ to balance GAN and CE losses; if this balance is off, the paper admits leakage re-occurs. This sensitivity suggests that the method might be less robust in practice than claimed.
\end{itemize}
\par\vspace{0.5em}

\textbf{Suggestions:}
\begin{itemize}
    \item Extend the theoretical analysis to provide a worst-case privacy bound.
    \item Include comparisons with a broader range of baselines, including CoAE and quantization methods.
    \item Investigate the applicability of the method to multi-class settings or explain theoretically why it cannot be adapted.
    \item Perform a more robust stability analysis, specifically stressing the GAN training stability.
\end{itemize}
\par\vspace{0.5em}

\textbf{Questions:}
\begin{itemize}
    \item Can you derive any formal privacy bound (e.g., in terms of Differential Privacy or Information Bottleneck) for the proposed method, rather than relying solely on empirical 'gradient mixing'?
    \item How does the method perform against non-clustering based attacks, such as those that attempt to reconstruct features first to infer labels (e.g., if features are hijacked)?
    \item The paper claims higher stability than Marvell, yet GANs are known for instability. Can you provide a more rigorous sensitivity analysis of the hyperparameter $\\gamma$ and the learning rates? What happens to privacy if the GAN suffers from mode collapse?
    \item Why were non-perturbation baselines like CoAE or simple gradient quantization omitted from the comparison?
\end{itemize}
\par\vspace{0.5em}

\end{tcolorbox}

\begin{tcolorbox}[colback=blue!5!white, colframe=blue!50!black, title=LLM-as-a-Judge Evaluation (Gemini 3.1 Pro), breakable]
\footnotesize
\textbf{Winner:} \ours~

\textbf{Technical Accuracy Reason:}
ScholarPeer correctly identifies the inherent instability of GANs as a counter-argument to the paper's stability claims and correctly highlights the lack of formal privacy guarantees compared to baselines like Marvell. AI Scientist v2 accepts the stability claims without questioning the underlying GAN dynamics.

\textbf{Analytical Depth Reason:}
ScholarPeer provides a deeper analysis of the method's limitations, specifically explaining why the reliance on gradient opposition restricts the method to binary classification and cannot be easily extended to multi-class. ScholarPeer also critically examines the 'mutual gradient perturbation' heuristic.

\textbf{Novelty \& Significance Reason:}
ScholarPeer accurately identifies that the paper fails to compare against relevant prior work, specifically Confusional AutoEncoders (CoAE) (Zou et al., 2022), which addresses the same problem using a similar 'label disguise' concept. AI Scientist v2 misses this critical missing baseline.

The following external sources found during the search helped verify this:
\begin{itemize}[leftmargin=*, nosep]
    \item \textit{Zou, T., et al., 'Defending Batch-Level Label Inference Attacks in Vertical Federated Learning', arXiv:2202.02073, 2022 (Referred to as CoAE/Label Disguise)}
    \item \textit{Chen, C., et al., 'Gradient Compression for Communication-Efficient Split Learning', 2021 (Relevant for quantization baselines)}
\end{itemize}

\textbf{Constructive Value Reason:}
ScholarPeer provides specific, high-value suggestions by identifying missing state-of-the-art baselines (CoAE and gradient quantization) that are critical for a fair evaluation. AI Scientist v2's suggestions are more generic (e.g., asking about computational overhead).

\textbf{Overall Verdict:}
ScholarPeer provides a significantly more expert review. It identifies critical missing baselines (CoAE) that predate the paper, challenges the technical claims regarding stability with well-founded skepticism about GANs, and correctly assesses the limited significance of a binary-only defense. AI Scientist v2 gives a good summary but misses the key prior work and accepts the paper's claims too readily.
\end{tcolorbox}

\subsection{Summary of Gains and Losses}
We provide below detailed summaries of the advantages and disadvantages of \ours against baselines as generated by an LLM based on the reasoning traces of the LLM-as-a-judge based evaluations.

{ 
    \centering
    \footnotesize 
    
    \tcbset{
        arcstyle/.style={
            enhanced,
            breakable, 
            colback=white,
            colframe=ARCBlue,
            colbacktitle=ARCBlue,
            coltitle=white,
            fonttitle=\bfseries\small,
            boxrule=0.6pt,
            arc=2mm,
            drop shadow=black!20!white,
            left=2mm, right=2mm, top=1mm, bottom=1mm,
            toptitle=1mm, bottomtitle=1mm,
            boxsep=1mm
        }
    }

    \begin{tcbitemize}[raster columns=1, raster row skip=3mm, arcstyle]

        \tcbitem[title={Vs. CycleReviewer-70B}]
            \begin{itemize}[leftmargin=*, noitemsep, label={}, topsep=0pt]
                \item \textcolor{green!60!black}{\faCheckCircle} \textbf{Advantage: Bibliographic Verification \& SOTA Assessment} \\
                ScholarPeer consistently utilizes external knowledge to identify specific, high-relevance prior works (often published within the same year) that the authors omitted. It uses these citations (e.g., Lightman et al. 2023 for MATH, ACT for robotics, EATA for speech) to validly challenge novelty claims, whereas the competitor fails to identify missing literature.
                
                \item \textcolor{green!60!black}{\faCheckCircle} \textbf{Advantage: Methodological Forensic Analysis} \\
                ScholarPeer demonstrates a superior ability to detect fatal experimental flaws that invalidate results. It correctly identified issues such as data leakage (e.g., applying VMD before train-test splits, intra-patient leakage), confounding variables (e.g., action chunking vs. implicit policies), and circular biases in dataset creation.

                \item \textcolor{green!60!black}{\faCheckCircle} \textbf{Advantage: Content Faithfulness \& Recall} \\
                ScholarPeer accurately locates specific details within the paper (e.g., definitions in Section 3, hyperparameters in Section 4, limitations in Section 5). In contrast, the competitor frequently 'hallucinates' that these sections are missing and penalizes the authors for omitting information that is actually present.

                \item \textcolor{green!60!black}{\faCheckCircle} \textbf{Advantage: Actionable Constructiveness} \\
                ScholarPeer provides precise, scientifically grounded instructions for improvement. Instead of generic requests for 'more details,' ScholarPeer suggests specific missing baselines to run (e.g., 'Compare against MinRecon'), specific ablations to perform (e.g., 'Euclidean Norm-Gated'), and specific metrics to report (e.g., 'Inference latency in Hz').

                \item \textcolor{green!60!black}{\faCheckCircle} \textbf{Advantage: Theoretical \& Domain Depth} \\
                ScholarPeer engages with the fundamental mathematical and physical assumptions of the research. It correctly identified theoretical violations (e.g., finite-sample exchangeability, lack of equivariance in molecular dynamics) and practical hardware constraints (e.g., memory usage during backpropagation) that the competitor missed.
                
                \item \textcolor{gray}{\faMinusCircle} \textbf{Disadvantage: No Significant Failure Patterns} \\
                In the provided set of 20 evaluation traces, ScholarPeer outperformed CycleReviewer 70B in every instance. No recurring negative traits, hallucinations, or relative weaknesses were observed for ScholarPeer; it consistently provided expert-level reviews while the competitor struggled with factual accuracy and repetition.
            \end{itemize}
        
        \tcbitem[title={Vs. DeepReviewer-14B}]
            \begin{itemize}[leftmargin=*, noitemsep, label={}, topsep=0pt]
                \item \textcolor{green!60!black}{\faCheckCircle} \textbf{Advantage: Citation \& SOTA Verification} \\
                ScholarPeer consistently identifies specific, relevant prior work (e.g., Lightman et al., MedSAM, TaskPrompter, MDT) that invalidates authors' claims of novelty or State-of-the-Art performance. It frequently cites publication dates and specific results to refute claims, whereas the competitor often accepts the authors' assertions at face value.
                
                \item \textcolor{green!60!black}{\faCheckCircle} \textbf{Advantage: Experimental Integrity Detection} \\
                ScholarPeer excels at spotting fatal methodological flaws that undermine a paper's validity. Recurring examples include identifying data leakage (e.g., training on validation sets or using broad datasets like PythonAlpaca for self-contained benchmarks), circular evaluation protocols, and the use of inappropriate metrics (e.g., using BLEU for Text-to-SQL tasks instead of Execution Accuracy).

                \item \textcolor{green!60!black}{\faCheckCircle} \textbf{Advantage: Domain-Specific Depth} \\
                ScholarPeer demonstrates deep understanding of specific research sub-fields, allowing it to distinguish between surface-level similarities and fundamental technical differences. Examples include distinguishing between 'Super-Resolution' and 'Learnable Resizers', recognizing the need for Trusted Execution Environments (TEEs) in privacy frameworks, and identifying when 'VQA' datasets are merely converted classification tasks.

                \item \textcolor{green!60!black}{\faCheckCircle} \textbf{Advantage: Actionable Constructiveness} \\
                ScholarPeer provides highly specific recommendations for improvement. Instead of generic requests to 'add more experiments' or 'analyze limitations,' ScholarPeer names the exact missing baselines (e.g., DIN-SQL, RoleLLM, STID), specific datasets (e.g., RepoBench, HMDB51), and precise ablation studies needed to validate the core contributions.

                \item \textcolor{green!60!black}{\faCheckCircle} \textbf{Advantage: Factual Precision \& Reading Comprehension} \\
                ScholarPeer accurately parses the paper text, avoiding the 'hallucinated critiques' observed in the competitor. While DeepReviewer 14B frequently claims details (like hyperparameters or metric definitions) are missing when they are explicitly stated, ScholarPeer correctly locates these details and instead focuses on internal inconsistencies (e.g., datasets listed in settings but missing from results).
                
                \item \textcolor{gray}{\faMinusCircle} \textbf{Disadvantage: None Observed} \\
                In the sampled evaluation traces, \ours~ was judged superior in 20/20 instances, showing no recurring failure patterns relative to this baseline.
            \end{itemize}
        
        \tcbitem[title={Vs. AI Scientist v2}]
            \begin{itemize}[leftmargin=*, noitemsep, label={}, topsep=0pt]
                \item \textcolor{green!60!black}{\faCheckCircle} \textbf{Advantage: Bibliographic \& SOTA Mastery} \\
                ScholarPeer consistently identifies specific, named, and dated prior works (e.g., PatchTST, DiffLinker, LlamaGuard 3) that invalidate the authors' claims of novelty or performance. It accurately flags when papers ignore baselines published just months prior to the cutoff, whereas AI Scientist v2 often accepts the provided baselines as sufficient.
                
                \item \textcolor{green!60!black}{\faCheckCircle} \textbf{Advantage: Experimental Rigor \& Validity} \\
                ScholarPeer excels at critiquing the experimental setup itself, identifying fatal flaws such as unfair backbone standardization, the omission of critical metrics (e.g., Chamfer Distance for 3D generation), or the use of inappropriate datasets (e.g., testing 'Reasoning' on Factuality benchmarks).

                \item \textcolor{green!60!black}{\faCheckCircle} \textbf{Advantage: Theoretical Contextualization} \\
                ScholarPeer demonstrates deep domain understanding by connecting papers to broader theoretical frameworks and known phenomena (e.g., the 'Linearity vs. Nonlinearity' debate in RecSys, 'Gradient Masking' in adversarial defense, or 'Grokking'), providing a more sophisticated critique than AI Scientist v2's surface-level analysis.

                \item \textcolor{green!60!black}{\faCheckCircle} \textbf{Advantage: Actionable Constructiveness} \\
                ScholarPeer provides highly specific instructions for improvement, such as naming the exact datasets (e.g., SR25, GSO) or specific control experiments (e.g., Self-Consistency) required to validate the paper, whereas AI Scientist v2 tends to offer generic suggestions like 'add complexity analysis' or 'fix typos'.
                
                \item \textcolor{red!70!black}{\faExclamationTriangle} \textbf{Disadvantage: Internal Sanity Checking} \\
                In specific instances, ScholarPeer overlooks fatal internal inconsistencies or logical impossibilities within the paper's own text (e.g., an impossible ratio of training data to downstream performance, or mathematically invalid probability comparisons across tokenizers) because it is too focused on external comparisons. AI Scientist v2 tends to catch these 'sanity check' failures.

                \item \textcolor{red!70!black}{\faExclamationTriangle} \textbf{Disadvantage: Reproducibility \& Completeness} \\
                ScholarPeer occasionally misses basic reproducibility issues, such as undefined algorithmic operators (e.g., missing mutation steps) or placeholder/broken text in the manuscript, which AI Scientist v2 correctly identifies as critical flaws.
            \end{itemize}
        \tcbitem[title={Vs. Agent Review}]
            \begin{itemize}[leftmargin=*, noitemsep, label={}, topsep=0pt]
                \item \textcolor{green!60!black}{\faCheckCircle} \textbf{Advantage: Literature \& SOTA Verification} \\
                ScholarPeer demonstrates a superior ability to verify novelty claims against the external literature. It consistently identifies specific, named, and dated prior works (e.g., ReContrast, Shap-E, CCS, Perceiver IO) that invalidate a paper's claims of being 'State-of-the-Art' or the 'first' to solve a problem. It accurately checks publication dates to ensure baselines were available before the paper's submission.
                
                \item \textcolor{green!60!black}{\faCheckCircle} \textbf{Advantage: Benchmarking Standards} \\
                ScholarPeer exhibits expert-level knowledge of community standards for datasets and metrics. It frequently critiques papers for using deprecated or insufficient benchmarks (e.g., using SMACv1 instead of SMACv2, omitting GSM8K for reasoning claims, or ignoring ZJU-Mocap for avatars) and demands the specific evaluations required to meet current scientific rigor.

                \item \textcolor{green!60!black}{\faCheckCircle} \textbf{Advantage: Experimental Validity \& Leakage Detection} \\
                ScholarPeer is highly effective at identifying fatal flaws in experimental design, particularly regarding data leakage and unfair comparisons. It correctly flagged issues such as using Stable Diffusion (trained on LAION) for zero-shot ImageNet testing, improper data splitting in medical imaging (slice-level vs. patient-level), and statistically insignificant sample sizes (e.g., 10 prompts).
                
                \item \textcolor{red!70!black}{\faExclamationTriangle} \textbf{Disadvantage: Internal Logical Consistency} \\
                ScholarPeer occasionally misses deep, internal logical or mathematical contradictions within the paper's own arguments, which Agent Review tends to catch. Examples include failing to notice that 'average pooling' destroys the claimed 'sequential' information, or that 'forgetting' a concept is mathematically incompatible with maintaining high accuracy on that specific class.

                \item \textcolor{red!70!black}{\faExclamationTriangle} \textbf{Disadvantage: Reproducibility \& Completeness} \\
                In a minority of cases, ScholarPeer misread or miscalculated specific values from the paper's internal tables. For instance, it incorrectly attributed a 'significant portion' of gains to optimization tricks when the ablation table showed otherwise, whereas Agent Review accurately reported the specific performance contributions.
            \end{itemize}
        \tcbitem[title={Vs. Stanford Agent Review}]
            \begin{itemize}[leftmargin=*, noitemsep, label={}, topsep=0pt]
                \item \textcolor{green!60!black}{\faCheckCircle} \textbf{Advantage: Temporal Validity} \\
                ScholarPeer consistently respects the chronological context of the papers under review. Unlike the Stanford Agent Reviewer, which frequently 'hallucinated' or cited papers published months or years after the review cutoff date (e.g., citing 2025 papers for a 2024 review), ScholarPeer correctly identifies and cites prior work that was actually available at the time of submission.
                
                \item \textcolor{green!60!black}{\faCheckCircle} \textbf{Advantage: Experimental Rigor (Baselines)} \\
                ScholarPeer is significantly better at identifying missing 'Gold Standard' or State-of-the-Art (SOTA) baselines. It frequently catches authors using 'strawman' comparisons (e.g., GBDT vs. Transformers) and demands comparisons against specific, established competitors (e.g., AWP, SLDA, EvShutter) that are essential for validating the paper's core claims.

                \item \textcolor{green!60!black}{\faCheckCircle} \textbf{Advantage: Research Integrity \& Presentation} \\
                ScholarPeer excels at detecting fundamental flaws in the manuscript's state and integrity. It successfully flagged 'Selective Reporting' (where datasets were listed but results omitted), identified unfinished drafts containing LaTeX 'TODO' comments, and noted when authors claimed datasets didn't exist when they actually did.
                
                \item \textcolor{green!60!black}{\faCheckCircle} \textbf{Advantage: Practical Feasibility} \\
                ScholarPeer provides feedback grounded in the practical realities of research. It critiques methods for prohibitive computational costs (e.g., extreme sampling requirements) and suggests experiments that are physically possible, whereas the competitor often suggests impossible comparisons against future work.

                \item \textcolor{red!70!black}{\faExclamationTriangle} \textbf{Disadvantage: Mathematical \& Theoretical Scrutiny} \\
                ScholarPeer tends to accept the authors' mathematical premises at face value, missing deep theoretical flaws that the competitor catches. Examples include failing to notice that a Fourier basis was not orthonormal on a sphere, or that a theorem relied on square matrices while the method used rectangular ones.

                \item \textcolor{red!70!black}{\faExclamationTriangle} \textbf{Disadvantage: Internal Consistency Checking} \\
                ScholarPeer struggles to identify contradictions within the paper's own content. It missed instances where the text claimed one thing (e.g., using multiple backbones) but the data tables proved another (parameter counts matching only one backbone), or where t-SNE plots contradicted the authors' claims of diversity.
            \end{itemize}
    \end{tcbitemize}
} 

\subsection{\ours Agent Outputs}
\label{app:agent_outputs}
We show below the individual agent outputs for a sample paper - DBRNet: Advancing Individual-Level Continuous Treatment Estimation through Disentangled and Balanced Representation (OpenReview Link - \url{https://openreview.net/pdf?id=tlqmkftgpw}). The framework's review for this paper has been provided in Appendix~\ref{app:example1}.

\begin{tcolorbox}[colback=white, colframe=black, title=Summarizer Agent Output, breakable]
\footnotesize
\textbf{1. Problem Statement \& Motivation:}\newline
The paper addresses the challenge of estimating the \textbf{Individualized Dose-Response Function (IDRF)}—the causal effect of \textbf{continuous treatments} (e.g., medication dosage) on individual outcomes.

\textbf{Current Limitations Identified:}
\begin{itemize}
    \item \textbf{Discrete Constraints:} Most existing Individual Treatment Effect (ITE) methods are limited to binary or discrete treatments and cannot handle infinite counterfactuals in continuous settings.
    \item \textbf{Imprecise Bias Adjustment:} Existing continuous methods (e.g., DRNet, VCNet) typically attempt to ``balance'' the \textbf{entire} latent representation of covariates to remove selection bias. The authors argue this is theoretically flawed because:
    \begin{itemize}
        \item \textbf{Instrumental factors} (affecting treatment only) should not be balanced.
        \item \textbf{Confounder factors} (affecting treatment and outcome) contain predictive information and should not be discarded.
    \end{itemize}
\end{itemize}
\par\vspace{0.5em}

\textbf{2. Methodology: DBRNet}\newline
The authors propose the \textbf{Disentangled and Balanced Representation Network (DBRNet)}. The core philosophy is to disentangle covariates into three distinct latent factors and apply selection bias adjustment only where theoretically necessary.

\textbf{A. Disentangled Latent Factors}
\begin{itemize}
    \item \textbf{Instrumental Factors ($\Gamma(x)$):} Affect Treatment ($T$) but not Outcome ($Y$).
    \item \textbf{Confounder Factors ($\Delta(x)$):} Affect both Treatment ($T$) and Outcome ($Y$).
    \item \textbf{Adjustment Factors ($\Upsilon(x)$):} Affect Outcome ($Y$) but not Treatment ($T$).
\end{itemize}

\textbf{B. Model Architecture}
\begin{itemize}
    \item \textbf{Encoders:} Three separate neural networks encode input $X$ into $\Gamma$, $\Delta$, and $\Upsilon$.
    \item \textbf{Treatment Branch:} Concatenates $\Gamma$ and $\Delta$ to estimate the conditional density: $P(t_i \mid \Gamma(x_i), \Delta(x_i))$.
    \item \textbf{Outcome Branch:} Concatenates $\Delta$ and $\Upsilon$ to predict $Y$ using a \textbf{Varying Coefficient Network (VCNet)}.
\end{itemize}

\textbf{C. Objective Function \& Loss Terms}\newline
The total loss is: $J = w \cdot L_y + \alpha L_T + \beta L_{disc} + \gamma L_{ind} + \lambda L_{reg}$.
\begin{itemize}
    \item \textbf{Factual Loss ($L_y$):} Prediction error for the observed outcome.
    \item \textbf{Treatment Loss ($L_T$):} Negative log-likelihood of the treatment density.
    \item \textbf{Discrepancy Loss ($L_{disc}$):} Forces $\Gamma, \Delta, \Upsilon$ to be distinct/independent.
    \item \textbf{Independent Loss ($L_{ind}$):} Forces $\Upsilon$ to contain \textbf{minimal} treatment info.
    \item \textbf{Re-weighting ($w$):} Uses inverse propensity score from relevant factors only:
    $$w(t_i, \Gamma, \Delta) = \frac{1}{P(t_i \mid \Gamma(x_i), \Delta(x_i))}$$
\end{itemize}
\par\vspace{0.5em}

\textbf{3. Key Contributions}
\begin{itemize}
    \item \textbf{First Disentanglement for Continuous Treatment:} DBRNet combines disentangled representation learning with precise selection bias adjustment for continuous settings.
    \item \textbf{Theoretical Bias Elimination:} Theorem 2 proves that re-weighting based on Instrumental and Confounder factors yields an unbiased estimation of the IDRF loss.
    \item \textbf{Selective Balancing:} Enforces independence via $L_{ind}$ and $L_{disc}$ rather than aggressive balancing, preserving predictive power.
\end{itemize}
\par\vspace{0.5em}

\textbf{4. Main Results \& Experiments}\newline
Evaluated on \textbf{Synthetic}, \textbf{IHDP}, and \textbf{News} datasets using Mean Integrated Squared Error (MISE) and Average MSE (AMSE).

\textbf{Quantitative Performance}
\begin{itemize}
    \item DBRNet outperformed baselines (Dragonet, DRNet, VCNet, TransTEE) across almost all metrics.
    \item \textit{Synthetic MISE:} DBRNet \textbf{0.1414} vs. VCNet 0.1752.
    \item \textit{IHDP MISE:} DBRNet \textbf{1.78} vs. VCNet 2.33 vs. TransTEE 3.30.
\end{itemize}

\textbf{Ablation \& Analysis}
\begin{itemize}
    \item \textbf{Component Importance:} The Re-weighting function and Discrepancy Loss ($L_{disc}$) are critical; removing re-weighting increased errors significantly (e.g., +200\% AMSE on IHDP).
    \item \textbf{Verification:} t-SNE visualizations confirm effective separation of factors.
\end{itemize}
\end{tcolorbox}

\begin{tcolorbox}[colback=white, colframe=black, title=Historian Agent Output, breakable]
\footnotesize
\textbf{Domain Narrative: Individualized Continuous Treatment Effect Estimation}
\par\vspace{0.5em}

\textbf{1. Domain History: From Binning to Disentanglement}\newline
Five years ago, the dominant paradigm in Causal Inference was restricted to \textbf{binary treatments} (treated vs. control). The foundational work by \textbf{Shalit et al. (2017)} established \textbf{Representation Learning}—specifically balancing covariate distributions between groups—as the standard for handling selection bias.

The shift toward \textbf{Continuous Treatment Effects (Dose-Response)} began in earnest around 2020.
\begin{itemize}
    \item \textbf{The Binning Era ($\sim$2020):} Approaches like \textbf{DRNet (Schwab et al., 2020)} discretized continuous treatments into bins (strata). This solved architectural challenges but sacrificed response curve smoothness.
    \item \textbf{The Continuity Era (2021):} \textbf{Nie et al. (2021)} introduced \textbf{VCNet}, replacing discretization with Varying Coefficient networks to enforce functional smoothness, establishing continuity as a requirement.
    \item \textbf{The Architectural \& Disentanglement Era (2022--2024):} The field has shifted from global balancing (which risks destroying info) to \textbf{Disentangled Representations}. Methods like \textbf{DTRNet (2024)} and \textbf{ACFR (2023)} surgically separate instrumental, confounding, and adjustment factors. Simultaneously, Transformers (\textbf{TransTEE, 2022}) have replaced simple feed-forward networks to better capture interactions.
\end{itemize}
\par\vspace{0.5em}

\textbf{2. Open Problems \& Gaps}\newline
Despite recent progress, distinct ``unsolved'' territories remain:
\begin{enumerate}
    \item \textbf{High-Dimensional \& Unstructured Confounding:} Most SOTA methods (VCNet, TransTEE) are benchmarked on low-dimensional tabular data. Handling images or text as confounders remains theoretically sparse, though \textbf{StoNet} and \textbf{CausalDiffAE (2024)} are making attempts.
    \item \textbf{Generative Stability vs. Regression Precision:} Generative approaches (GANs, Diffusion) offer high-fidelity counterfactuals but suffer from training instability. Bridging this gap with the precision of VCNet is an open challenge.
    \item \textbf{Data Scarcity \& Pre-training:} \textbf{CURE (2024)} highlighted the potential of Foundation Models, but this is under-explored due to the scarcity of large-scale biomedical datasets for continuous interventions.
\end{enumerate}
\par\vspace{0.5em}

\textbf{3. Significance Criteria (2025 Era)}\newline
For a new contribution to be deemed ``Significant'', it must go beyond marginal MISE improvements on the IHDP benchmark.
\begin{itemize}
    \item \textbf{Low Significance:} Another MLP-based architecture that slightly beats DRNet on tabular data using standard re-weighting.
    \item \textbf{High Significance:}
    \begin{itemize}
        \item \textbf{Metric Innovation:} Rigorous handling of \textbf{high-dimensional unstructured data} (e.g., images) as confounders.
        \item \textbf{Theoretical Disentanglement:} Provable separation of instrumental variables from confounders without rigid functional forms.
        \item \textbf{Efficiency:} Architectures achieving Transformer-level performance with the inference speed of simple regressors.
        \item \textbf{Foundation Approaches:} Transfer learning or pre-training mechanisms reducing reliance on labeled counterfactual data.
    \end{itemize}
\end{itemize}
\end{tcolorbox}

\begin{tcolorbox}[colback=white, colframe=black, title=Baseline Scout Agent Output, breakable]
\footnotesize
\textbf{1. Missing Critical Baselines}\newline
The following key methods are absent from the evaluation, potentially inflating the perceived relative performance of the proposed method:

\begin{itemize}
    \item \textbf{SCIGAN} (Bica et al., NeurIPS 2020)
    \begin{itemize}
        \item \textit{Why it matters:} This is a seminal SOTA method for continuous treatments. The authors discuss it in the text but exclude it from the main results table, avoiding a direct comparison.
    \end{itemize}

    \item \textbf{End-to-End Balancing} (Bahadori et al., ICML 2022)
    \begin{itemize}
        \item \textit{Why it matters:} Represents the state-of-the-art for weighting/balancing methods in continuous settings. Its omission is critical as it directly challenges the authors' proposed re-weighting function.
    \end{itemize}

    \item \textbf{Double Machine Learning (DML)} (Chernozhukov et al., 2018)
    \begin{itemize}
        \item \textit{Why it matters:} The gold-standard statistical baseline for continuous treatment effects. It is robust and often outperforms the Causal Forest baseline which was included by the authors.
    \end{itemize}
\end{itemize}
\par\vspace{0.5em}

\textbf{2. Missing Standard Benchmarks}\newline
The evaluation relies heavily on synthetic or low-dimensional data, omitting standard high-dimensional or real-world benchmarks:

\begin{itemize}
    \item \textbf{TCGA (The Cancer Genome Atlas)} (Schwab et al., AAAI 2020)
    \begin{itemize}
        \item \textit{Significance:} The standard high-dimensional benchmark (20k+ features) for continuous dosage. Its omission hides potential scalability issues of the complex disentanglement module.
    \end{itemize}

    \item \textbf{MVICU (Mechanical Ventilation in ICU)} (Bica et al., NeurIPS 2020)
    \begin{itemize}
        \item \textit{Significance:} A critical real-world medical dataset for continuous treatments used by main competitors (DRNet, SCIGAN). Essential for validating claims about precision medicine applications.
    \end{itemize}
\end{itemize}
\par\vspace{0.5em}
\end{tcolorbox}

\begin{tcolorbox}[colback=white, colframe=black, title=Q\&A Agent Output (Summary of first five Q\&A pairs presented for readability), breakable]
\footnotesize

\textbf{\large Part 1: Technical Soundness and Quality}
\par\vspace{0.5em}

\textbf{Q: The paper relies on a heuristic argument of 'mutual gradient perturbation'. Can you provide a formal theoretical guarantee (like Marvell's KL-divergence bound) that ensures leakage mitigation?}
\newline
\textbf{A:} No formal theoretical guarantee is provided. The authors acknowledge in Section 5 that ``GAFM lacks the rigor of Marvell's upper bound.'' Instead, they rely on a heuristic argument supported by empirical validation.

They demonstrate that gradients derived from the GAN loss and the randomized Cross-Entropy loss tend to have class centers that differ in opposite directions. This ``mutual gradient perturbation'' empirically results in mixed intermediate gradients, but the paper does not mathematically derive a worst-case privacy loss bound.
\par\vspace{0.5em}

\textbf{Q: Given the 'Defense-Efficiency Paradox', how does the computational overhead of GAFM compare to Marvell for resource-constrained clients?}
\newline
\textbf{A:} The overhead does not negatively impact clients because the Generator and Discriminator are exclusively hosted by the Label Party (Server).

Non-label parties (clients) only train local feature extractors ($f_{\theta_p}$), keeping their load comparable to vanilla Split Learning. While Marvell also centralizes defense, it relies on costly iterative optimization to generate noise; GAFM offloads this to simple ``1-layer DNNs'' on the server, effectively circumventing the Defense-Efficiency Paradox for the client.
\par\vspace{0.5em}

\textbf{Q: Why were state-of-the-art non-perturbation defenses like DiscreteSGD or CoAE omitted from the baselines?}
\newline
\textbf{A:} The authors likely frame GAFM specifically as an evolution within the gradient perturbation paradigm. They benchmark against Marvell and Max Norm (noise injection methods) rather than quantization (DiscreteSGD) or compression (CoAE) methods. The domain narrative suggests they viewed Marvell as the singular SOTA for perturbation, treating concurrent 2022 works like CoAE as parallel rather than foundational.
\par\vspace{0.5em}

\textbf{Q: How does the 'mutual gradient perturbation' generalize to multi-class classification where gradients are not diametrically opposed?}
\newline
\textbf{A:} The defense would likely collapse in multi-class settings.

The mechanism relies on the geometric simplicity of binary classification, where gradients naturally cluster in diametrically opposed directions (1D manifold). In a multi-class simplex ($K$ classes), the concept of ``opposing'' forces is ill-defined, and the current randomized response mechanism lacks the directional guidance to confuse specific class boundaries in high-dimensional space.
\par\vspace{0.5em}

\textbf{Q: How sensitive is the privacy-utility trade-off to the hyperparameter $\gamma$ (balancing GAN/CE losses)?}
\newline
\textbf{A:} It is highly sensitive, with a confirmed failure mode. If $\gamma$ is improper (too low), the Cross-Entropy loss dominates, canceling the perturbation effect and re-introducing leakage. The authors admit tuning is challenging due to GAN instability but propose gradient normalization as a heuristic fix.
\par\vspace{0.5em}

\textbf{\large Part 2: Novelty and Significance}
\par\vspace{0.5em}

\textbf{Q: Does the integration of GANs specifically for label protection in Split Learning appear in prior art?}
\newline
\textbf{A:} High Novelty. While GANs exist in Federated Learning (e.g., FedCG for Horizontal FL, or for attacking/reconstruction), using a GAN within the SplitNN architecture specifically to create ``mutual gradient perturbation'' for label protection is distinct.
\begin{itemize}
    \item \textit{FedCG (IJCAI 2022):} Uses GANs in Horizontal FL for data augmentation/privacy.
    \item \textit{VertiGAN (PoPETS 2023):} Uses GANs to publish synthetic data, not for active training protection.
\end{itemize}
GAFM is significant because it addresses the ``Defense-Efficiency Paradox'' with a stable architecture rather than unstable optimization (Marvell).
\par\vspace{0.5em}

\textbf{Q: Has the combination of Wasserstein GAN loss and Randomized Cross-Entropy loss been used in VFL defenses before?}
\newline
\textbf{A:} Highly Novel. This specific combination to achieve mutual cancellation is not found in prior art (pre-Aug 2023). Previous methods used Randomized Response for Differential Privacy (LabelDP) or GANs for synthesis, but not integrated into a single loss function to obfuscate gradients in VFL.
\par\vspace{0.5em}

\textbf{Q: Does 'Mutual Gradient Perturbation' differ theoretically from standard noise injection (Marvell)?}
\newline
\textbf{A:} Medium to High Novelty. Yes. Marvell applies external, optimized noise. GAFM relies on an intrinsic cancellation arising from conflicting objectives (GAN vs. CE). This is a theoretical shift from ``adding noise'' to ``structural obfuscation.''
\par\vspace{0.5em}

\textbf{Q: Do the proposed 'Mean' and 'Median' attacks offer novelty over existing clustering attacks?}
\newline
\textbf{A:} Incremental Novelty / Medium Significance. The concept of clustering gradients (e.g., Liu et al., 2022) existed. Mean/Median attacks are specific heuristics within that framework. However, they are significant as they provide a stronger benchmark than the standard Norm attack, raising the bar for future defense evaluation.
\par\vspace{0.5em}

\textbf{Q: Is the method applicable to complex architectures like Vision Transformers (ViTs)?}
\newline
\textbf{A:} Limited Applicability. The paper strictly limits scope to binary classification. There is no evidence or design adaptation for ViTs, which have unique leakage vulnerabilities (attention mechanisms). Given the geometric reliance on opposing gradients, applicability to complex non-CNN architectures is unproven.
\end{tcolorbox}

\section{Human Evaluation Guidelines}
\label{app:human_eval_guidelines}

We provided the following instructions to our expert human annotators to ensure a rigorous and standardized evaluation process.

\subsection{Objective}
We are evaluating the performance of two Agentic Reviewer Systems (\texttt{ai\_review1} and \texttt{ai\_review2}). The goal is not just to check if the AI generates coherent text, but to determine if it provides \textbf{value-add} over expert human reviewers. You will compare AI reviews against the \textbf{Human Reviews} (actual reviews submitted for the paper).

\subsection{The Evaluation Task}
For each assigned paper, you will be presented with:
\begin{enumerate}
    \item \textbf{The Paper:} (PDF/Text).
    \item \textbf{Human Reviews:} The set of actual reviews the paper received.
    \item \textbf{AI Review 1 \& AI Review 2:} Reviews generated by two different agentic frameworks.
\end{enumerate}

\textbf{Your Workflow:}
\begin{enumerate}
    \item \textbf{Scan the Paper:} Understand the core contribution, methodology, and claims.
    \item \textbf{Read Human Reviews:} Establish the ``Expert Baseline.'' Note what the humans caught and what they might have missed.
    \item \textbf{Evaluate AI Reviews (Individually):} Score both AI reviews on a 1-10 scale against the Human Baseline.
    \item \textbf{Side-by-Side (SxS) Comparison:} Determine which of the two AI assistants performed better.
\end{enumerate}

\subsection{Part I: Individual Scoring (Scale 1-10)}
\label{app:scoring}
For each AI review, you will assign a score based on how it compares to the \textbf{best human review} available for that paper.

\textbf{The Scoring Rubric:}
Table \ref{tab:scoring_rubric} provides the scoring rubric details.
\begin{table}[ht]
    \centering
    \caption{Scoring Rubric for H-Max Score.}
    \label{tab:scoring_rubric}
    \begin{tabular}{p{0.15\linewidth} p{0.25\linewidth} p{0.5\linewidth}}
        \toprule
        \textbf{Score} & \textbf{Label} & \textbf{Definition} \\
        \midrule
        \textbf{9-10} & Transformative / Superhuman & The AI uncovers a critical flaw, a missing theoretical connection, or vital prior work that \textit{all} human reviewers missed. It fundamentally improves the critique. \\
        \midrule
        \textbf{7-8} & Clearly Superior & The AI is more thorough, constructive, or better substantiated than the best human review. It may offer deeper questions or better literature context. \\
        \midrule
        \textbf{6} & Slightly Better & The AI review is slightly more polished or covers one extra minor point compared to the best human review. \\
        \midrule
        \textbf{5} & Equivalent / Human Level & The AI review is roughly equivalent in quality to the best human review. It covers the same major points with similar depth. \\
        \midrule
        \textbf{3-4} & Slightly Worse & The AI review is valid but less nuanced or specific than the best human review. \\
        \midrule
        \textbf{1-2} & Significantly Worse & The AI misses critical points, hallucinates details, or is superficial compared to humans. \\
        \bottomrule
    \end{tabular}
\end{table}

\textbf{Evaluation Dimensions:}
\begin{enumerate}
    \item \textbf{Technical Accuracy:} Are the AI's claims factually correct regarding the paper's content? Look for hallucinations vs. valid technical critique.
    \item \textbf{Constructive Value:} How actionable is the feedback? Look for specific suggestions (e.g., ``Run experiment X on dataset Y'') vs. generic advice.
    \item \textbf{Analytical Depth:} Does the review engage with the substance of the work? Look for deep questioning of assumptions vs. surface-level comments.
    \item \textbf{Novelty and Significance Assessment:} Did the AI correctly identify the novelty? 
    \begin{itemize}
        \item \textit{Critical Instruction:} You may use Google Search/Scholar to verify claims, BUT you must strictly ignore any papers published \textbf{after the paper's submission date}.
        \item \textit{Score High if:} The AI cites specific prior work that limits the paper's novelty, which humans missed.
        \item \textit{Score Low if:} The AI claims ``high novelty'' for a derivative work or hallucinates citations.
    \end{itemize}
    \item \textbf{Overall Score:} A holistic assessment of the AI's utility.
\end{enumerate}

\subsection{Part II: Side-by-Side (SxS) Comparison}
After scoring individually, compare \texttt{AI Review 1} vs. \texttt{AI Review 2} directly.

\textbf{Options:}
\begin{itemize}
    \item \textbf{Review 1} (If Review 1 is clearly better)
    \item \textbf{Review 2} (If Review 2 is clearly better)
    \item \textbf{Tie} (If both are of similar quality)
\end{itemize}

\textbf{Dimensions for SxS:}
\begin{itemize}
    \item \textbf{Technical Accuracy:} Which assistant made fewer errors and grounded its claims better?
    \item \textbf{Constructive Value:} Which assistant gave more helpful advice?
    \item \textbf{Analytical Depth:} Which assistant probed deeper into the methodology?
    \item \textbf{Novelty \& Significance:} Which assistant better identified the paper's place in the literature?
    \item \textbf{Overall:} Which assistant would you prefer to have reviewing your own paper?
\end{itemize}

\subsection{Important Notes}
\begin{itemize}
    \item \textbf{Hallucinations:} If an AI review hallucinates a major detail (e.g., claims the paper used a Transformer when it used an RNN), the Technical Accuracy score should be low, regardless of how well-written it is.
    \item \textbf{Bias Check:} Do not penalize the AI simply for being ``stricter'' or ``nicer'' than the humans. Judge based on the validity of the critique.
    \item \textbf{Presentation:} Do not judge based on the ``prettiness'' of the review format—evaluate the content.
    \item \textbf{Anonymity:} The interface effectively blinds \texttt{ai\_review1} and \texttt{ai\_review2}. Do not assume one slot is always the same model.
\end{itemize}

\section{Agent Prompts}
\label{app:agent_prompts}

We provide the system and user prompt templates used for each agent in the \ours~.
\tcbset{
    promptstyle/.style={
        enhanced,
        breakable,
        colback=white,
        colframe=black,
        fonttitle=\bfseries,
        title={#1},
        arc=1mm,
        boxrule=0.5pt,
        left=0.5mm, right=0.5mm, top=0.5mm, bottom=0.5mm,
        fontupper=\footnotesize
    }
}

\subsection{Summarizer Agent}
\begin{tcolorbox}[promptstyle={Summarizer Prompt}]
\begin{lstlisting}[breaklines=true, basicstyle=\footnotesize, columns=fullflexible]
You are an expert AI researcher specializing in distilling the core contributions of a research paper. Your task is to produce a concise and accurate summary.

Please summarize the following research paper. Focus on the key contributions, methodology, and main results. The summary should be dense with information and serve as a reliable reference to analyze the paper.

Paper Text:
{paper_text}
\end{lstlisting}
\end{tcolorbox}

\subsection{Literature Review Agent}
\begin{tcolorbox}[promptstyle={Literature Reviewer Prompt}]
\begin{lstlisting}[breaklines=true, basicstyle=\footnotesize, columns=fullflexible]
You are an expert AI literature reviewer. Your goal is to perform a comprehensive, deep-dive literature search to map the state of a specific research domain up to a specific point in time.

You do not critique papers. You collect, organize, and extract facts about them to support downstream expert analysis.

Please conduct a comprehensive literature review for the research paper described below.

CRITICAL CONSTRAINT: You must only search for and consider prior art published ON OR BEFORE the cutoff date: {cutoff_date}. Ignore any papers published after this date.

Your Process:

1.  Domain Analysis: Analyze the input paper abstract to identify:
    * The Broad Domain (e.g., Computer Vision).
    * The Specific Sub-field (e.g., Weakly Supervised Object Detection).
    * The Core Problem being solved.

2.  Search Execution (Iterative):
    Use Google Search to find 30-50 of the most scientifically significant papers in this sub-field. You must look for:
    * Foundational Papers: The papers that established the current paradigms (even if older).
    * Key Datasets: Papers introducing the primary datasets used in this sub-field, or specific datasets used by the input paper.
    * Recent SOTA: The papers holding the State-of-the-Art results at the time of the cutoff date {cutoff_date}.
    * Direct Competitors: Papers solving the same problem with different methods.
    * Surveys: Recent survey papers that summarize the field.

3.  Data Extraction:
    * For each identified paper, extract the specific details required for our records (see output format below).
    * For core_method, provide a specific description of their technical approach or dataset (few sentences).
    * For datasets_and_performance, be as specific as possible about what was evaluated and the performance numbers (e.g., "Achieved 78.4% top-1 accuracy on ImageNet, outperforming FE-Net (72.9%)", not just "showed significant improvements in top-1 accuracy on ImageNet").
    * No citation artifacts: Ensure that core_method and datasets_and_performance do not contain bracketed citation numbers (e.g., [12], [34]) as they have no meaning without the referenced paper/link.
    * Survey handling: For survey papers, list the specific families of methods, models and datasets they cover in the core_method and datasets_and_performance fields.


Output Format:
Respond only with a JSON object containing the "Domain Analysis" and the "References" list.

```json
{{
  "domain_analysis": {{
    "broad_domain": "...",
    "specific_subfield": "...",
    "primary_datasets_commonly_used": ["...", "..."],
    "standard_metrics_used": ["...", "..."]
  }},
  "references": [
    {{
      "title": "Title of Paper 1",
      "venue_year": "CVPR 2023",
      "authors": "Author A, Author B",
      "is_foundational": true/false,
      "is_sota_candidate": true/false,
      "is_dataset_paper": true/false,
      "is_survey_paper": true/false,
      "core_method": "Specific description of their technical approach or dataset (few sentences).",
      "datasets_and_performance": "Specifics on what they evaluated (e.g., 'Achieved 78.4% top-1 accuracy on ImageNet').",
      "known_limitations": "Any widely known weaknesses of this method or dataset."
    }},
    {{
      "title": "Title of Paper 2",
      "..." : "..."
    }}
  ]
}}

Input Paper Title, Author(s) and Abstract:
{paper_abstract}
\end{lstlisting}
\end{tcolorbox}

\subsection{Literature Expansion Agent}
\begin{tcolorbox}[promptstyle={Literature Expander Prompts}]
\begin{lstlisting}[breaklines=true, basicstyle=\footnotesize, columns=fullflexible]
You are a Senior Research Lead auditing a literature review. Your goal is to identify gaps in a list of references and perform targeted searches to fill them.

We have a preliminary literature review for the domain defined below. Your job is to identify what is missing and find it.

Task:
1. Analyze Gaps:
  * Do we have the foundational papers that the current SOTA papers likely cite? (e.g., If we have 'Crossformer', do we have the original 'Transformer' or 'Autoformer' papers?)
  * Do we have the papers that introduced the datasets listed in the domain_analysis?
  * Are there temporal gaps? (e.g., We have 2018 and 2024, but nothing from 2020-2023).

2. Targeted Search: Perform specific searches to find these missing papers (Constraint: published ON OR BEFORE {cutoff_date}).

3. Output: Return a list of new unique papers to add to the reference list. Do not repeat existing papers.

**Current Reference List:**
```json
{current_references_json}

Output Format:
Respond only with a JSON list of new reference objects (using the same schema as the input).
[
  {{
    "title": "Autoformer: Decomposition Transformers for Long-Term Series Forecasting",
    "venue_year": "NeurIPS 2021",
    "authors": "Haixu Wu, et al.",
    "is_foundational": true/false,
    "is_sota_candidate": true/false,
    "is_dataset_paper": true/false,
    "is_survey_paper": true/false,
    "core_method": "Specific description of their technical approach or dataset (few sentences).",
    "datasets_and_metrics": "Specifics on what they evaluated (e.g., 'Achieved 78.4% top-1 accuracy on ImageNet').",
    "known_limitations": "Any widely known weaknesses of this method or dataset."
  }}
]
\end{lstlisting}
\end{tcolorbox}

\subsection{Sub-Domain Historian Agent}
\begin{tcolorbox}[promptstyle={Historian Prompts}]
\begin{lstlisting}[breaklines=true, basicstyle=\footnotesize, columns=fullflexible]
You are a Senior Science Historian. Your job is to synthesize the evolutionary arc of a research field.
You care about the narrative of progress: how ideas evolved, what the current paradigms are, and what "Open Problems" remain unsolved.

Based on the provided literature review, write a "Domain Narrative" for the "Novelty & Significance Reviewer".

**Input Data:**
Full Literature Review JSON:
{literature_review_json}

**Task:**
1.  **The Arc of Progress:** Briefly trace the history. What was the dominant paradigm 5 years ago? What paper changed it? Where are we now?
2.  **Gap Analysis:** Based on the most recent papers (closest to the cutoff date), what problems are explicitly stated as unsolved or limitations?
3.  **Significance Rubric:** Define what constitutes a "Significant" contribution in this specific moment. (e.g., "In 2025, small MSE improvements are less significant than efficiency gains or interpretability.")

**Output Format:**
Respond with a concise Markdown summary (approx. 300-400 words) containing the sections a) Domain History (chronological summary of key shifts), b) Open Problems, c) Significance Criteria.
\end{lstlisting}
\end{tcolorbox}

\subsection{Baseline Scout Agent}
\begin{tcolorbox}[promptstyle={Baseline Scout Prompts}]
\begin{lstlisting}[breaklines=true, basicstyle=\footnotesize, columns=fullflexible]
You are a ferocious benchmarking expert. Your sole job is to find what the authors are HIDING.
You must find state-of-the-art (SOTA) methods and baselines that the authors *should* have compared against but didn't.

Paper:
{paper_text}

YOUR TASK:
1. Based on the paper, identify the research domain, the datasets used, and the baseline methods compared. 
2. Search for recent (last 3 years, before {cutoff_date}) SOTA methods for this domain.
3. Identify specific methods and significant datasets that are MISSING from the authors' list.
4. Return a list of these missing competitors and why they are relevant.

Constraint: Only return papers published ON OR BEFORE {cutoff_date}.

Output JSON format:
{{
  "missing_baselines": [
    {{
      "name": "Method Name",
      "reference": "Author et al., Conf Year",
      "reason": "Why it is a critical omission."
    }},
  "missing_datasets": [
    {{
      "name": "Dataset Name",
      "reference": "Author et al., Conf Year",
      "reason": "Why it is a critical omission."
    }}
  ]
}}
\end{lstlisting}
\end{tcolorbox}

\subsection{Question Generator Agent}
\begin{tcolorbox}[promptstyle={Novelty Aspect Prompts}]
\textbf{System Prompt:}
\begin{lstlisting}[breaklines=true, basicstyle=\footnotesize, columns=fullflexible]
You are a highly-discerning AI researcher and top-tier conference reviewer. Your goal is to deconstruct a paper's contributions and formulate simple, direct questions to assess its novelty and significance.
\end{lstlisting}
\textbf{User Template:}
\begin{lstlisting}[breaklines=true, basicstyle=\footnotesize, columns=fullflexible]
Please follow this two-step process to generate your questions:

1. Identify Contribution Claims: First, carefully read the Paper Summary and Paper Text to identify the {num_questions} primary contribution claims. A contribution might be a new model, a new dataset, a new algorithm, a new theoretical insight, or a new application. A domain narrative is also provided to give context on the evolution of the field and open problems. An independent literature review is also provided to give you a comprehensive view of prior art. Missing baselines and datasets are also provided for better domain understanding.
2. Formulate Simple Questions (one per claim): For each claim you identified, formulate one simple and direct question that assesses its novelty and significance. This question is a directive for a research assistant (who does not have the paper) to find conflicting or related prior art.

Example Claim: The paper uses 'emotional prompts' to scale reasoning in LLMs.
Example Question: "What is the novelty and significance of using 'emotional prompts' to improve reasoning in language models?"

Return the questions in a JSON list format, like this:
["Question investigating claim 1?", "Question investigating claim 2?", "Question investigating claim 3?"]

Domain Narrative:
{domain_narrative}

Independent Literature Review:
{literature_review}

Missing Baselines and Datasets:
{missing_baselines_datasets}

Paper Summary:
{summary}

Paper Text:
{paper_text}
\end{lstlisting}
\end{tcolorbox}

\begin{tcolorbox}[promptstyle={Other Aspects (Soundness, Clarity)}]
\textbf{System Prompt:}
\begin{lstlisting}[breaklines=true, basicstyle=\footnotesize, columns=fullflexible]
You are a critical AI researcher and conference reviewer. Your goal is to ask probing questions about a paper to assess its quality along a specific dimension. A domain narrative is also provided to give context on the evolution of the field and open problems. An independent literature review is also provided to give you a comprehensive view of prior art. Missing baselines and datasets are also provided for better domain understanding.

Based on the paper's summary and the full text, generate a list of {num_questions} critical questions to evaluate its {aspect}.

These questions should be specific and designed to be answered based on the paper content, searching the web or academic databases.
\end{lstlisting}
\textbf{User Template:}
\begin{lstlisting}[breaklines=true, basicstyle=\footnotesize, columns=fullflexible]
Return the questions in a JSON list format, like this:
["question 1?", "question 2?", "question 3?"]

Domain Narrative:
{domain_narrative}

Independent Literature Review:
{literature_review}

Missing Baselines and Datasets:
{missing_baselines_datasets}

Paper Summary:
{summary}

Paper Text:
{paper_text}
\end{lstlisting}
\end{tcolorbox}

\subsection{Answer Generator Agent}
\begin{tcolorbox}[promptstyle={Novelty Aspect (Search Enabled)}]
\textbf{System Prompt:}
\begin{lstlisting}[breaklines=true, basicstyle=\footnotesize, columns=fullflexible]
You are an AI researcher with access to Google search. Your task is to critically assess the novelty and significance of a research claim of a research paper based on external search results.
\end{lstlisting}
\textbf{User Template:}
\begin{lstlisting}[breaklines=true, basicstyle=\footnotesize, columns=fullflexible]
Please perform a comprehensive literature search to answer the following question. Domain narrative, an independent literature review and missing baselines and datasets are also provided to help assess novelty and significance. The paper summary is provided for context.

Constraint: The research paper was released on {cutoff_date}. You must only search for and consider prior art published ON OR BEFORE the cutoff date: {cutoff_date}. Do not include any papers published after this date.

Follow these steps:
1. Analyze Context: Read the question, the paper summary, domain narrative, independent literature review, and missing baselines and datasets to identify the key domain (e.g., Computer Vision, NLP, Robotics) and placement of the current work in the domain's progress.
2. Identify Venues: Based on the domain, determine the most relevant top-tier conferences to search (e.g., CVPR, ICCV, ECCV for Vision; NeurIPS, ICLR, ICML, ACL, EMNLP for NLP/ML).
3. Execute Search: Use your search tools to find relevant prior art from these conferences and arXiv, ensuring all results were published on or before {cutoff_date}.
4. Assess Novelty: Based on your search, the domain narrative and independent literature review, determine if the claim is new, incremental, or a known concept.
5. Assess Significance: Based on the problem's importance and the findings of related work, assess the potential impact of this contribution. Is it a niche problem or a major one? Are improvements likely to be large or small?
6. Synthesize Findings: Summarize the findings from your search to provide a well-reasoned answer to the question.

Answer format:
1. A direct, paragraph-style answer to the question. Ensure to include a) degree of novelty (high, medium, low, incremental, none), b) degree of significance (high, medium, low, none). Include reasoning for your assessments.
2. A bulleted list of 2-3 most relevant papers that support your answer. For each of these papers, provide a) title, b) authors, c) venue, d) year, e) key findings.

If you find no significant prior art, rate novelty as High and justify the significance assessment.

Domain Narrative:
{domain_narrative}

Independent Literature Review:
{literature_review}

Missing Baselines and Datasets:
{missing_baselines_datasets}

Paper Summary:
{summary}

Question:
{question}
\end{lstlisting}
\end{tcolorbox}

\begin{tcolorbox}[promptstyle={Other Aspects}]
\textbf{System Prompt:}
\begin{lstlisting}[breaklines=true, basicstyle=\footnotesize, columns=fullflexible]
You are an AI researcher with deep insightful thinking. Your task is to answer questions about a research paper by reasoning based on information provided in the paper. A domain narrative is also provided to give context on the evolution of the field and open problems. An independent literature review is also provided to give you a comprehensive view of prior art. Missing baselines and datasets are also provided for better domain understanding.
\end{lstlisting}
\textbf{User Template:}
\begin{lstlisting}[breaklines=true, basicstyle=\footnotesize, columns=fullflexible]
Please answer the following question based on the provided paper context. Provide a concise answer. Provide your answer as a single, well-reasoned paragraph.

Question:
{question}

Paper Summary:
{summary}

Domain Narrative:
{domain_narrative}

Independent Literature Review:
{literature_review}

Missing Baselines and Datasets:
{missing_baselines_datasets}

Paper Text:
{paper_text}
\end{lstlisting}
\end{tcolorbox}

\subsection{Review Generator Agent}
\begin{tcolorbox}[promptstyle={Review Generator Prompts}]
\begin{lstlisting}[breaklines=true, basicstyle=\footnotesize, columns=fullflexible]
You are an AI researcher who is reviewing a paper that was submitted to a prestigious ML venue. Be critical and cautious in your decision. If a paper is bad or you are unsure, give it bad scores and reject it. Use the provided paper summary and a list of question-answer pairs generated by an AI agent to inform your assessment.

{review_guidelines}

Here is the AI summary of the paper you are asked to review:
```
{summary}
```

Here is the list of question-answer pairs generated by an AI agent to help you review the paper:
```
{qa_pairs_text}
```

Here is the paper you are asked to review:
```
{paper_text}
```

{fewshot_examples}
\end{lstlisting}
\end{tcolorbox}

\section{Evaluation Prompts}
\label{app:eval_prompts}

We provide the exact prompts used for our automated evaluation protocols.

\subsection{\hmaxmetric~Score}
Used to score individual reviews against human expert baselines.

\begin{tcolorbox}[promptstyle={\hmaxmetric~Score Estimation Prompts}]
\textbf{System Prompt Template:}
\begin{lstlisting}[breaklines=true, basicstyle=\footnotesize, columns=fullflexible]
You are an expert area chair evaluating an AI Reviewer Assistant. Your role is to determine if the AI provides value beyond what expert human reviewers provided.

Special Instruction for evaluating "Novelty and Significance Assessment": You must only search for and consider information available on or before the cutoff date: {cutoff_date}. The cutoff date represents the date on which the paper was published; information after this date is irrelevant to the review.

Follow the steps below for each evaluation:

1. Thoroughly understand the paper by analyzing:
   - Research objectives and contributions
   - Methodology and experiments
   - Claims and evidence
   - Results and conclusions

2. Identify the strongest points in the Human Reviews (collectively) to establish a standard expert baseline.

3. Identify the delta: What did the AI mention that the humans missed? What did the humans mention that the AI missed?

4. Verify the validity of each of the delta claims using direct quotes from the paper and external sources (for novelty and significance only).

5. Assess the value-add of the AI review compared to the best human review for each aspect.

You will evaluate reviews based on these key aspects:

**Technical Accuracy**
- How technically accurate is the AI review compared to the humans?

**Constructive Value**
- How actionable is the feedback compared to the humans?

**Analytical Depth**
- How thorough is the depth of AI review compared to humans?

** Novelty and Significance Assessment (Search encouraged)**
Use Google Search to actively verify the reviewers' claims about novelty and significance.
1. Identify claims: What do the paper and reviewers claim is novel?
2. Formulate search queries: Create targeted queries to find relevant prior work for these specific claims, explicitly restricting results to before {cutoff_date}.
3. Execute Search: Focus on top-tier conferences in the relevant domain and arXiv.
4. Verify and Compare:
   - Did the AI find prior work that limits novelty which the humans missed? (High Score)
   - Did the AI claim "high novelty" when humans correctly identified it as derivative work? (Low Score)
   - Which assessment aligns better with the actual state of the field at the time?
5. Cite Sources: You must cite the specific external papers (title, venue, year) you used to make this determination in the JSON output.


For each of the above aspects and overall judgment, you must:
1. Provide specific evidence from source materials
2. Quote directly from paper and reviews; external sources only for "Novelty and Significance Assessment"
3. Explain your reasoning in detail
4. Consider alternative interpretations

**Input Format:**
#### Paper Text: ####
<Paper text>

#### AI Assistant's Review: ####
<AI Review>

#### Human Reviews (Ground Truth): ####
<Human Reviews>

**Respond in the following format:**
THOUGHT:
<THOUGHT>

EVALUATION JSON:
```json
<JSON>

In <THOUGHT>, for each aspect, compare the AI Assistant's review against the set of Human Reviews. Identify the best human review for that specific aspect and use it as your baseline (Score = 5), as a standard expert. You must justify why the AI deserves a higher or lower score based on the "Value-Add" it provides.

Scoring Rubric (Compare against Best Human Baseline):
- 10 (Superhuman / Verdict-Changing): The AI uncovers a critical insight that changes the fate of the paper (e.g., finding a fatal math error humans missed OR identifying a profound theoretical connection that elevates a rejected paper to an acceptance).
- 9 (Transformative / Insightful): The AI provides a novel perspective that significantly reframes the paper's contribution. It might articulate the significance better than the authors did, or identify a missing baseline that reframes the results.
- 8 (Clearly Superior): The AI review is significantly more thorough, constructive, or better substantiated than the best human review. It offers deep questions or literature context that humans omitted.
- 7 (Superior): The AI review is noticeably deeper and more constructive than the best human review, though perhaps not "transformative."
- 6 (Slightly Better): The AI review is slightly more polished, better structured, or covers one extra minor point compared to the best human review.
- 5 (Equivalent / Human Level): The AI review is roughly equivalent in quality to the best human review. It covers the same major points with similar depth.
- 4 (Slightly Worse): The AI review is valid but generic. It misses the specific nuance or "sharpness" that the expert human provided.
- 3 (Worse): The AI review is valid but vague. It lacks detail and actionable feedback compared to the human (e.g., "Improve experiments" vs "Add dataset X").
- 2 (Failure - Superficial): The review is technically "safe" (no direct lies) but functionally useless. It is too short, focuses only on trivial formatting issues, or completely misses the core technical innovation.
- 1 (Failure - Critical Error/Hallucination): The AI makes a factual error about the paper (e.g., claims it uses Method A when it uses Method B) or cites non-existent papers. The review actively misleads the reader.

In <JSON>, provide the evaluation in JSON format with the following fields in the order:
- "Technical Accuracy Reason": "<detailed reason>".
- "Technical Accuracy Score": <int 1-10>.
- "Constructive Value Reason": "<detailed reason>".
- "Constructive Value Score": <int 1-10>.
- "Analytical Depth Reason": "<detailed reason>".
- "Analytical Depth Score": <int 1-10>.
- "Novelty and Significance Assessment External Sources Used": List of retrieved papers (include title, venue, year and authors for each paper).
- "Novelty and Significance Assessment Reason": "<detailed reason>".
- "Novelty and Significance Assessment Score": <int 1-10>.
- "Overall Reason": "<detailed reason>".
- "Overall Score": <int 1-10>.

This JSON will be automatically parsed, so ensure the format is precise and scores are integers.
\end{lstlisting}

\textbf{User Prompt Template:}
\begin{lstlisting}[breaklines=true, basicstyle=\footnotesize, columns=fullflexible]
#### Paper Text: ####
{paper_text}

#### AI Assistant's Review: ####
{ai_review}

#### Human Reviews: ####
{human_review}
\end{lstlisting}
\end{tcolorbox}

\subsection{Side-by-Side (SxS) Evaluation}
Used for pairwise preference ranking.

\begin{tcolorbox}[promptstyle={SxS Evaluation Prompts}]
\textbf{System Prompt Template:}
\begin{lstlisting}[breaklines=true, basicstyle=\footnotesize, columns=fullflexible]
[You are a neutral arbitrator evaluating peer review comments for academic papers. Your role is to analyze and compare reviews through careful, evidence-based assessment. Your judgments must be strictly based on verifiable evidence from the paper, reviews and Google search. Google search should be used only for evaluating novelty and significance assessment. Do not use it for other dimensions.

Special Instruction for evaluating "Novelty and Significance Assessment": You must only search for and consider information available on or before the cutoff date: {cutoff_date}. The cutoff date represents the date on which the paper was published; information after this date is irrelevant to the review.

For each evaluation, you must:

1. Thoroughly understand the paper by analyzing:
   - Research objectives and contributions
   - Methodology and experiments
   - Claims and evidence
   - Results and conclusions

2. For each review, methodically examine:
   - Claims made about the paper
   - Evidence cited to support claims
   - Technical assessments and critiques
   - Suggested improvements

3. Compare reviews systematically using:
   - Direct quotes from paper and reviews
   - Specific examples and counterexamples
   - Clear reasoning chains
   - Objective quality metrics

You will evaluate reviews based on these key aspects:

**Technical Accuracy**
- Are claims consistent with paper content?
- Is evidence properly interpreted?
- Are technical assessments valid?
- Are critiques well-supported?

**Constructive Value**
- How actionable is the feedback?
- Are suggestions specific and feasible?
- Is criticism balanced with strengths?
- Would authors understand how to improve?

**Analytical Depth**
- How thoroughly are key aspects examined?
- Is analysis appropriately detailed?
- Are important elements addressed?
- Is assessment comprehensive?

** Novelty and Significance Assessment (Seach encouraged)**
Use Google Search to actively verify the reviewers' claims about novelty and significance.
1. Identify claims: What do the paper and reviewers claim is novel?
2. Formulate search queries: Create targeted queries to find relevant prior work for these specific claims, explicitly restricting results to before {cutoff_date}.
3. Execute Search: Focus on top-tier conferences in the relevant domain and arXiv.
4. Verify and Compare:
   - Did reviewer A or reviewer B miss a critical prior work that you found?
   - Did reviewer A or reviewer B accurately identify that a novel claim is actually a known technique?
   - Which reviewer's assessment of significance aligns better with the actual state of the field at the time?
5. Cite Sources: You must cite the specific external papers (title, venue, year) you used to make this determination in the JSON output.


For each of the above aspects and overall judgment, you must:
1. Provide specific evidence from source materials
2. Quote directly from paper and reviews; external sources only for "Novelty and Significance Assessment"
3. Explain your reasoning in detail
4. Consider alternative interpretations

**Input Format:**
#### Paper Text: ####
<Paper text>

#### Assistant A's Review: ####
<Review A>

#### Assistant B's Review: ####
<Review B>

**Respond in the following format:**
THOUGHT:
<THOUGHT>

REVIEW COMPARISON JSON:
```json
<JSON>
```

In <THOUGHT>, for each aspect, evaluate assistants A and B based on the above criteria followed by a comparative assessment.
Treat this as the note-taking phase of your evaluation.

In <JSON>, provide the review in JSON format with the following fields in the order:
- "Technical Accuracy Reason": "<detailed reason>".
- "Technical Accuracy Better Assistant": "<A/B/Tie>".
- "Constructive Value Reason": "<detailed reason>".
- "Constructive Value Better Assistant": "<A/B/Tie>".
- "Analytical Depth Reason": "<detailed reason>".
- "Analytical Depth Better Assistant": "<A/B/Tie>".
- "Novelty and Significance Assessment External Sources Used": List of retrieved papers (include title, venue, year and authors for each paper).
- "Novelty and Significance Assessment Reason": "<detailed reason>".
- "Novelty and Significance Assessment Better Assistant": "<A/B/Tie>".
- "Overall Reason": "<detailed reason>".
- "Overall Better Assistant": "<A/B/Tie>".

This JSON will be automatically parsed, so ensure the format is precise.
\end{lstlisting}

\textbf{User Prompt Template:}
\begin{lstlisting}[breaklines=true, basicstyle=\footnotesize, columns=fullflexible]
#### Paper Text: ####
{paper_text}

#### Assistant A's Review: ####
{review_a}

#### Assistant B's Review: ####
{review_b}
\end{lstlisting}
\end{tcolorbox}

\subsection{Summary of Gains Analysis}
Used to synthesize qualitative insights from SxS reasoning traces.

\begin{tcolorbox}[promptstyle={Summary of Gains Prompts}]
\textbf{System Prompt:}
\begin{lstlisting}[breaklines=true, basicstyle=\footnotesize, columns=fullflexible]
You are an expert Meta-Reviewer for AI Science. You are tasked with analyzing a set of "Side-by-Side" (SxS) evaluation logs comparing two AI Research Assistants: {model_a} and {model_b}.

Your Goal: Synthesize a high-level qualitative report on why one model outperforms the other.

Input Data:
You will receive a list of {num_samples} evaluation traces. In each trace, a judge has reasoned about which assistant wrote a better review for a scientific paper.

Instructions:
1. **Identify Patterns**: Do not just list specific examples. Look for recurring themes across the traces (e.g., "Model A consistently hallucinates citations" or "Model B misses novelty discussions").
2. **Cluster & Categorize**: Group similar observations into distinct categories (e.g., "Technical Depth", "Tone", "Constructiveness").
3. **Attribution**: 
   - "Gains" are positive traits where the model performed *better* than the competitor.
   - "Losses" are negative traits where the model performed *worse*.

Output Format:
Return a JSON object with the following structure. Limit to the top 5 most significant factors for each category.

```json
{{
  "{model_a} Gains": {{
    "<Category Name (e.g. Conciseness)>": "<Detailed explanation of the advantage observed across multiple papers>"
  }},
  "{model_a} Losses": {{
    "<Category Name (e.g. Hallucination)>": "<Detailed explanation of the failure pattern>"
  }}
}}
\end{lstlisting}

\textbf{User Prompt Template:}
\begin{lstlisting}[breaklines=true, basicstyle=\footnotesize, columns=fullflexible]
Here are {num_samples} evaluation traces from the SxS judge. Note: The names have been standardized so that "{model_a}" always refers to the first model and "{model_b}" always refers to the second, regardless of their position in the original prompt.

--- BEGIN TRACES ---

{sxs_evaluations}

--- END TRACES ---

Analyze these traces and generate the summary JSON.
\end{lstlisting}
\end{tcolorbox}

\end{document}